%% file: main.tex
\documentclass{vgtc}

\usepackage{mathptmx}
\usepackage{graphicx}
\usepackage{times}
\usepackage{comment}
\usepackage{subfigure} 
\usepackage{etoolbox}%
\usepackage{algpseudocode}
\usepackage{algorithm}
\usepackage{enumitem}

\usepackage{subfiles}
\usepackage{graphicx}
\usepackage{subfigure}
\usepackage{epstopdf}
\usepackage{comment}
\usepackage{float}
\usepackage[mathscr]{euscript}
\usepackage{etoolbox}
\usepackage{wrapfig}
\usepackage{booktabs}
\usepackage{hhline}

\usepackage[skip=8pt]{caption}

\usepackage{amsmath,amsthm,amssymb,latexsym,amscd}

\graphicspath{{./figs/}}

\newcommand{\Rspace}        {{\mathbb R}}
\newcommand{\Xspace}        {{\mathbb X}}

\newcommand {\mm}[1] {\ifmmode{#1}\else{\mbox{\(#1\)}}\fi}

\newcommand{\Hgroup}        {{\sf H}}

\newcommand{\denselist}{\vspace{-5pt} \itemsep -2pt\parsep=-1pt\partopsep -2pt}

\newcommand{\filtration} {{F}}
\newcommand{\PD} {\mm{PD}}

\newcommand{\Rips}{{R}}

\usepackage[bookmarks,backref=true,linkcolor=black]{hyperref} 
\hypersetup{
  pdfauthor = {},
  pdftitle = {},
  pdfsubject = {},
  pdfkeywords = {},
  colorlinks=true,
  linkcolor= black,
  citecolor= black,
  pageanchor=true,
  urlcolor = black,
  plainpages = false,
  linktocpage
}

\ifpdf
  \pdfoutput=1\relax                   
  \usepackage{graphicx}                
  \DeclareGraphicsExtensions{.pdf,.png,.jpg,.jpeg} 
\else
  \usepackage{graphicx}                
  \DeclareGraphicsExtensions{.eps}     
\fi%

\graphicspath{{figures/}{pictures/}{images/}{./}} 

\usepackage{microtype}                 
\PassOptionsToPackage{warn}{textcomp}  
\usepackage{textcomp}                  
\usepackage{mathptmx}                  
\usepackage{times}                     
\usepackage{cite}                      
\usepackage{tabu}                      
\usepackage{booktabs}                  

\onlineid{0}

\vgtccategory{Research}

\vgtcinsertpkg

\title{Visual Detection of Structural Changes in Time-Varying Graphs Using Persistent Homology}

\author{Mustafa Hajij\thanks{e-mail: mhajij@usf.edu}\\ %
        \scriptsize University of South Florida %
\and Bei Wang\thanks{e-mail: beiwang@sci.utah.edu}\\ %
     \scriptsize University of Utah %
\and Carlos Scheidegger\thanks{e-mail: cscheid@email.arizona.edu}\\ %
     \scriptsize University of Arizona %
\and Paul Rosen\thanks{e-mail: prosen@usf.edu}\\ %
     \scriptsize University of South Florida}

\shortauthortitle{Hajij \MakeLowercase{\textit{et al.}}: Find Structural Changes in Graphs Using Persistent Homology}


\teaser{
\centering
\includegraphics[width=0.78\linewidth]{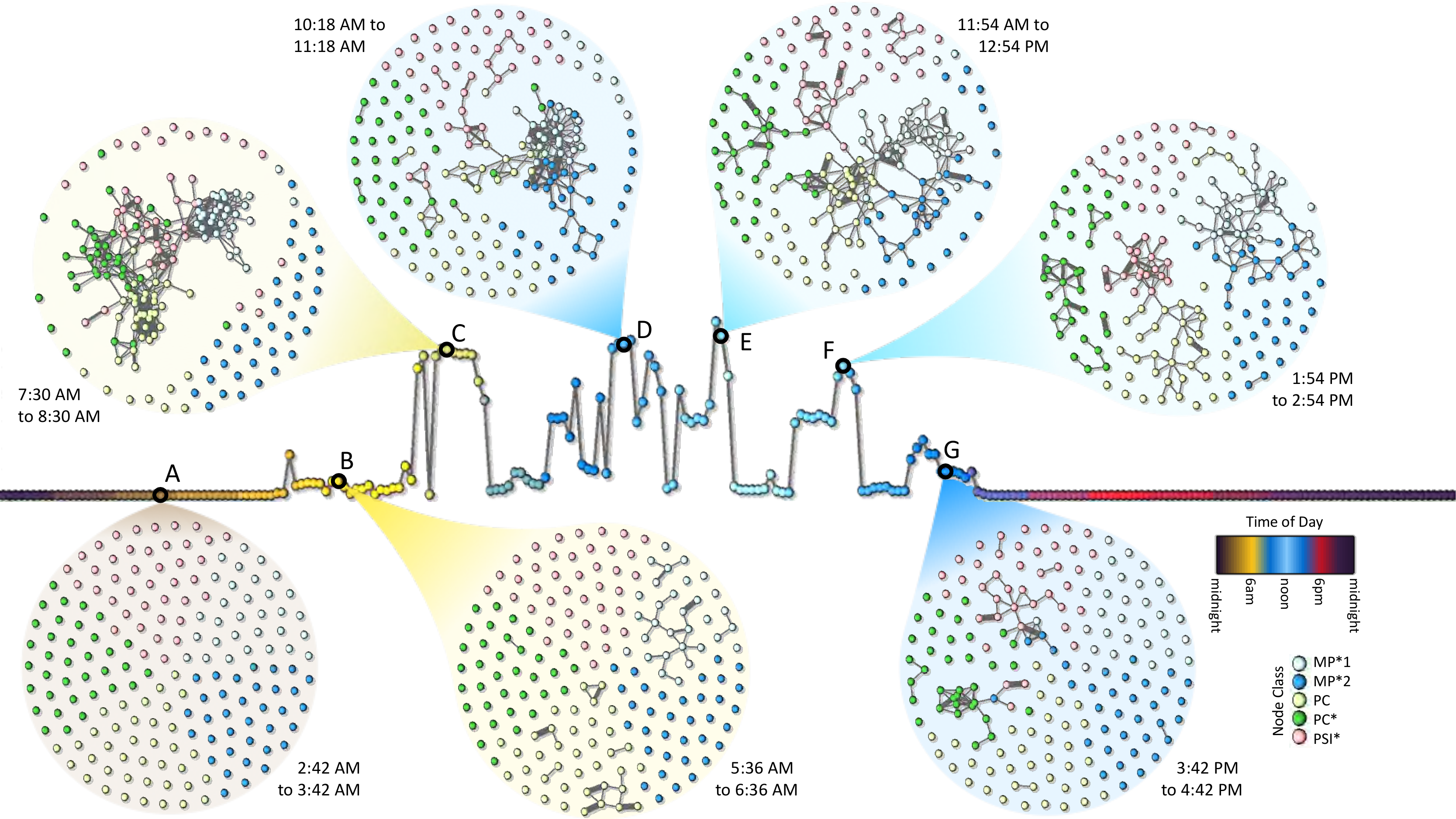}

\caption{Timeline showing the first Monday of the High School Communication Network dataset. The timeline is generated by comparing the commute-time $0$-dimensional homological features of the time-varying network using the bottleneck distance. 
Here, the $0$-dimensional homological features capture cluster-like behaviors in the data at multiple scales. The timeline differentiates periods of highly connected behaviors, such as instances C, D, E, and F, from periods of low or no activity, such as A, B, or G.}
\label{fig.teaser}
}

\input{sec-abstract}

\begin{document}

\maketitle

\input{sec-intro}

\input{sec-motivation}
\input{sec-priorwork}

\input{sec-methods}

\input{sec-evaluation}

\input{sec-discussion-intuitiveness}

\input{sec-discussion-stability}

\input{sec-conclusion}

\bibliographystyle{abbrv}

\bibliography{network_tda_vis}
\end{document}

%% file: sec-abstract.tex
\abstract{
Topological data analysis is an emerging area in exploratory data analysis and data mining. Its main tool, persistent homology, has become a popular technique to study the structure of complex, high-dimensional data. 
In this paper, we propose a novel method using persistent homology to quantify structural changes in time-varying graphs.  
Specifically, we transform each instance of the time-varying graph into metric spaces, extract topological features using persistent homology, and compare those features over time.  
We provide a visualization that assists in time-varying graph exploration and helps to identify patterns of behavior within the data. 
To validate our approach, we conduct several case studies on real world data sets and show how our method can find cyclic patterns, deviations from those patterns, and one-time events in time-varying graphs. We also examine whether persistence-based similarity measure as a graph metric satisfies a set of well-established, desirable properties for graph metrics. 
}

%% file: sec-intro.tex
\section{Introduction}
\label{intro}

Time-varying graphs are ubiquitous across many disciplines, yet difficult to analyze, making them a natural target for visualization -- a good visual representation of a time-varying graph will present its \emph{structure} and \emph{structural changes} quickly and clearly, to enable further analysis and exploration. 

A major development in graph drawing has been the observation that using \emph{derived} information can retain structure in static graph visualizations. For example, the dot layout uses \emph{node ranks} to perform hierarchical drawings~\cite{GansnerKoutsofiosNorth1993}; the neato algorithm employs \emph{graph distances} within statistical multidimensional scaling~\cite{GansnerKorenNorth2005}; Noack's energy model utilizes \emph{approximated clustering}~\cite{Noack2007}. 

In this paper, we take the first steps towards using topological features -- captured by persistent homology -- with the design goal of \emph{detecting potentially important structural changes in time-varying graph} data. By topological features, we do not mean the configuration of nodes and edges alone, but instead the $0$- and $1$-dimensional homology groups of a metric space that describe its connected components and tunnels, respectively.

This definition allows us to quantify structural elements within time-varying graphs to identify behavior patterns in the data. 
Persistent homology quantifies individual topological features (events) in the graph according to their significance (or \emph{persistence}). The set of all features, encoded by the \emph{persistence diagram}, can be seen as a fingerprint for the graph. Using this fingerprint, the most topologically-important structures of two graphs can be compared in a manner that is robust to small perturbations in the data.

Well-understood techniques in topological data analysis typically focus on the qualitative study of point cloud data under the metric space setting. In order to study graph data, our approach is to embed the graph in a metric space, where topological techniques can be applied. In other words, the notion of metric space acts as an organizational principle~\cite{Carlsson2014} in interpreting the graph data. 

\begin{figure*}[!t]
\centering
\includegraphics[width=0.82\linewidth]{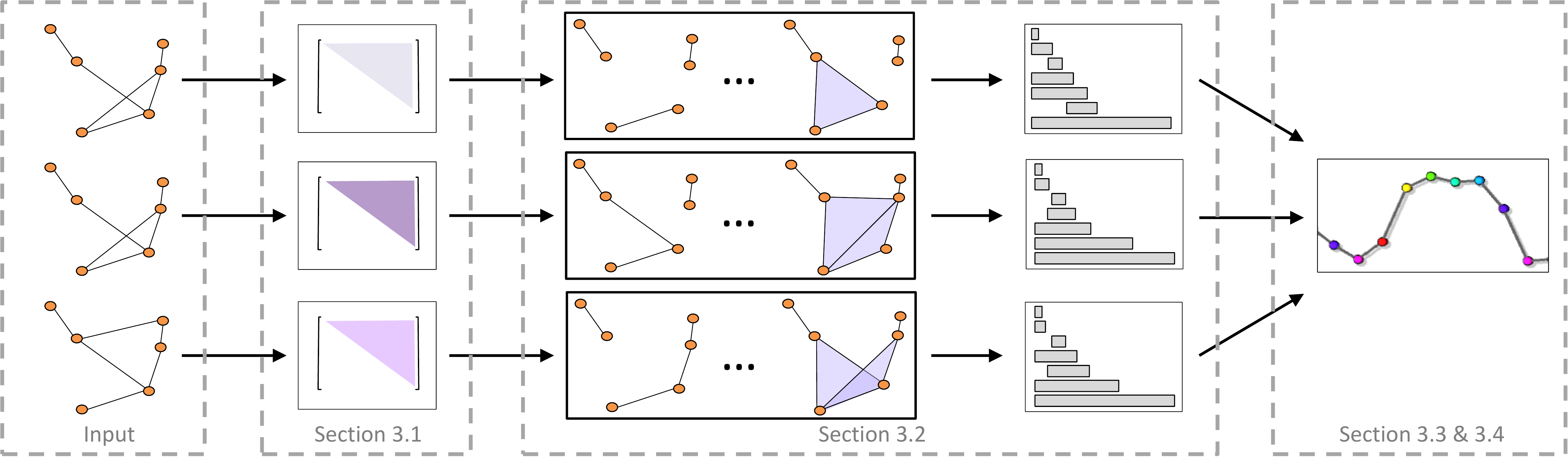}
\caption{The pipeline of our approach. 
An ordered sequence of graphs representing a time-varying graph is given as an input.
Each graph instance is individually embedded into a metric space (Section~\ref{sec.approach.metric}). 
The topological features of each (metric-space-embedded) graph instance is extracted, by computing persistent homology of its corresponding Rips filtration; the topological features are encoded by persistence diagrams and visualized as barcodes (Section~\ref{sec.approach.topology}). 
Finally, persistence diagrams are compared and the structural changes among the graph instances are visualized (Section~\ref{sec.approach.compare} and~\ref{sec.approach.vis}).}
\label{fig.pipeline}
\end{figure*}

Our approach, as seen Figure~\ref{fig.pipeline}, can be summarized as follows. The input of our pipeline is a time-varying graph, which is an ordered sequence of graph \textit{instances}. First, each instance is embedded into a metric space. Second, topological features of each instance are extracted using persistent homology, and encoded within persistence diagrams. Third, instances are compared by calculating the distance between persistence diagrams and projecting them using classical multidimensional scaling (MDS)~\cite{borg2005modern}. 

The data is then visualized using an interactive timeline and node-link diagrams, as shown in Figure~\ref{fig.teaser}. In this figure, the horizontal axis is used to represent time, while the vertical location is the first component of MDS, in other words, it captures the dissimilarities among instances. Graph instances from selected timeframes are drawn using a force-directed layout to demonstrate how the approach highlights different structure in the graph. 
The contributions of our paper are:
\begin{itemize}\denselist
\item A novel pipeline for detecting structural changes in time-varying graphs that uses persistent homology to summarize important structures, as opposed to directly comparing nodes and edges.
\item An interface that uses conventional visualization approaches adapted to the design goal of highlighting structural changes.
\item Two case studies of time-varying graphs showing how our approach can find cyclic patterns, deviations from those patterns, and unique one-time events in the graphs.
\item A study of the suitability of using persistence-based similarity measure for detecting structural changes in time-varying graphs. 
\end{itemize}

%% file: sec-priorwork.tex
\section{Related Work}

\paragraph{Static Graph Analysis and Visualization.}
We provide a brief overview here. See von Landesberger et al.'s survey~\cite{vonLandesbergerKujiperSchreck2011} for a full treatment. 

The first automated technique for node-link diagrams is Tutte's barycentric coordinate embedding~\cite{Tutte1963}, followed by linear programming techniques~\cite{GansnerKoutsofiosNorth1993}, force-directed/mass-spring embeddings~\cite{FruchtermanReingold1991, Hu2005}, embeddings of the graph metric~\cite{GansnerKorenNorth2005}, and linear-algebraic properties of the connectivity structures (especially, the graph Laplacian and associated eigenspaces)~\cite{KhouryHuKrishnanScheidegger2012, KorenCarmenHarel2002}.

Most graph visualization systems, including Gephi~\cite{bastian2009gephi}, NodeXL~\cite{HansenShneidermanSmith2010}, and Graphviz~\cite{ellson2002graphviz}, use variations on node-link visualizations to display graphs. For dense graphs, edge bundling can reduce visual clutter by routing graph edges to the same portion of the screen~\cite{HoltenVanWijk2009}. In terms of quality, divided edge bundling~\cite{SelassieHellerHeer2011} produces high-quality results, while hierarchical edge bundling~\cite{GansnerHuNorthScheidegger2011} scales to millions of edges with slightly lower quality. Because these quality and runtime trade-offs are so characteristic of node-link diagram visualizations, whether or not this class of diagrams can effectively unlock the insights hidden inside the structure of large networks remains an open research question. 

Other visual metaphors have been proposed to reduce clutter, ranging from relatively conservative proposals~\cite{DunneShneiderman2013, DwyerRicheMarriotMears2013} to variants of matrix diagrams~\cite{DinklaWestenbergWijk2012} and abstract displays of graph statistics~\cite{KairamMacLeanSavvaHeer2012}.

\paragraph{Time-Varying Graph Analysis.}
The problem we address is closely related to the problem of measuring  similarity or dissimilarly between graphs without knowing node correspondences. 
Comparing between graphs up to isomorphism is hard~\cite{Babai2016}. For this reason many notions of graph similarities have been proposed~\cite{baur2005network,  papadimitriou2010web}. These methods rely on mapping the graphs into a feature space and then defining distances on that space. Other approaches use kernel functions to build a similarity measures on graphs~\cite{ li2011graph, vishwanathan2010graph}. While large portions of the literature on graph similarity focus on graph comparison with known node correspondences, there are attempts to tackle the problem where node correspondence is unknown~\cite{vishwanathan2010graph, vogelstein2011shuffled}. Distance functions on the space of graphs have also been studied~\cite{chartrand1985edge}. 

\paragraph{Time-Varying Graph Visualization.}
Beck et al.~\cite{beck2014state} provide a detailed survey of dynamic graph visualization. They divide the techniques into two major categories, animation and timelines. Our approach falls into the latter category. Animation approaches, such as the work of Misue et al.~\cite{misue1995layout}, vary the graph representation over time, while making the graph as legible as possible at any given instance. Timeline approaches, such as the work of Greilich et al.~\cite{greilich2009visualizing}, use a non-animated, often spatially-oriented, visual channel to show the changes in the graph over time. Timeline approaches seem to provide a better overview of the data as it tries to capture the entire graph sequence in a single image. These approaches include multiple techniques such as node-link-based methods~ \cite{javed2012exploring},  matrix-based approaches \cite{burch2013matrix} and feature vector-based method~\cite{van2016reducing}. For more references and background see also Landesberger et al.'s survey~\cite{vonLandesbergerKujiperSchreck2011}.

\paragraph{Topological Data Analysis of Networks.}
Persistent homology is becoming an emerging tool in studying complex networks~\cite{ ELuYao2012, DonatoPetriScolamiero2012} including collaboration~\cite{CarstensHoradam2013, BampasidouGentimis2014} and  brain networks \cite{ DabaghianMemoliFrank2012, CassidyRaeSolo2015}.
To the best of our knowledge, our approach is the first in connecting  topological technique with the visualization design of (time-varying) graphs.

%% file: sec-methods.tex
\section{Approach}
\label{sec.approach}

Our approach uses persistent homology to identify and compare features in a time-varying graph. Our visual design goal is to identify  high-level structural changes in the graph. 
To do this, consider a time-varying graph $\mathcal{G}=\{G_0,...,G_n\}$, which contains an ordered sequence of static graph \textit{instances} $G_i = (V_i,E_i)$. 

We are interested in quantifying and visualizing structural changes of $\mathcal{G}$.
Our analysis pipeline (see Figure~\ref{fig.pipeline}) is described below, and we provide detailed description of each step in the subsequent sections. 
\begin{enumerate}\denselist

	\item Associate each instance $G_i$ with a metric space representation. 
This yields a symmetric distance matrix $d_{i}$, where $d_{i}(x, y)$ measures the (shortest-path or commute-time) distance between vertices $x$ and $y$ in $G_i$ (Section~\ref{sec.approach.metric}).   

	\item Extract topological features of $G_i$ by constructing a filtration $\filtration_i$ from its distance matrix $d_{i}$ and computing its corresponding $p$-dimensional persistence diagrams $\PD_p(\filtration_i)$ for $p \in \{0, 1\}$ (Section~\ref{sec.approach.topology}).  
    
	\item Capture the structural differences between $G_i$ and $G_j$ by computing the bottleneck or Wasserstein distance between their corresponding persistence diagrams $\PD_p(\filtration_i)$ and $\PD_p(\filtration_j)$ (Section~\ref{sec.approach.compare}). 
    
	\item Visualize the structural differences among the instances of $\mathcal{G}$ (Section~\ref{sec.approach.vis}). 
    
\end{enumerate}

\input{sec-methods-metric}

\input{sec-methods-topo}

\input{sec-methods-pd}

\input{sec-methods-vis}

\input{sec-methods-example}

%% file: sec-methods-metric.tex
\subsection{Graphs and Metric Space Representations}
\label{sec.approach.metric}

Suppose an instance $G_i$ is represented as a weighted, undirected graph with a vertex set $V$ and an edge set $E$ equipped with a positive edge weight $w$. We associate each graph instance $G_i$  with a metric space representation, which yields a symmetric distance matrix $d_i$. 

Consider the positive edge weight as the \emph{length} of an edge, then a natural metric $d_{sp}$ is obtained on $G_i$, where for every pair of vertices $x$ and $y$ in $G_i$, the distance $d_{sp}(x,y)$ is the length of the shortest-path between them. 
This is the classic \emph{shortest-path distance}, which is typically computed with Dijkstra's algorithm~\cite{Dijkstra1959} and its variations.

Alternatively, other distance metrics based on the graph Laplacian \cite{cvetkovic1980spectra}, such as commute-time distance, discrete biharmonic distance and diffusion distance, can be considered. For instance, the \emph{commute-time distance} is defined as~\cite{FoussPirotteRenders2007},
\begin{equation}
\label{commute}
d_{ct}^2(x,y)=\sum_{i=1}^{|V|-1} \frac{1}{\lambda_i} (\phi_i(x)-\phi_i(y))^2. 
\end{equation}
Here $\{\lambda_i\}_{i=0}^{|V|-1}$ and $\{\phi\}_{i=0}^{|V|-1}$ are the generalized eigenvalues and eigenvectors of the graph Laplacian of $G_i$, respectively~\cite{chung1997spectral}. 
In practice, we approximate the summations of Equation (\ref{commute})  by considering the first few eigenvectors, since the higher eigenvectors do not contribute significantly. 

\begin{figure}[h]
	\centering
	\subfigure[\small{Shortest-path Distance}\label{fig.shortestpath.distance.mesh}]{
		\hspace{5pt}
        \includegraphics[width=0.30\linewidth]{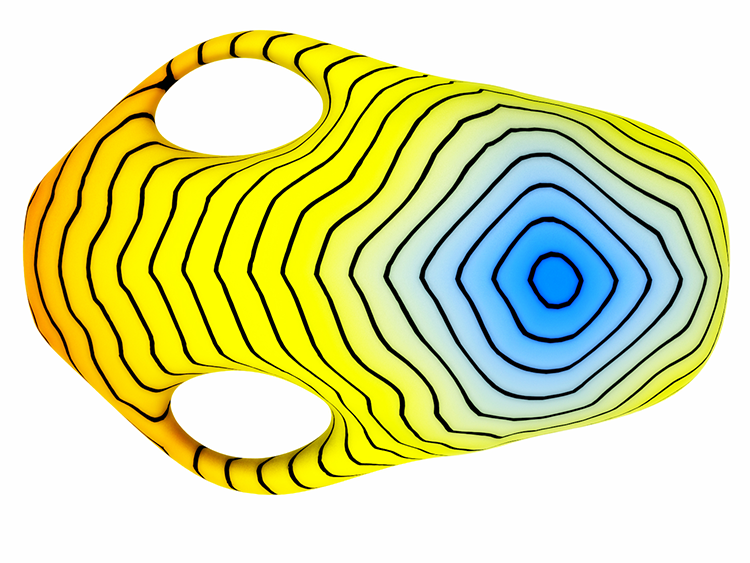}
        \hspace{5pt}
    }
    \hspace{15pt}
	\subfigure[\small{Commute-time Distance}\label{fig.commute.distance.mesh}]{
    	\hspace{5pt}
		\includegraphics[width=0.30\linewidth]{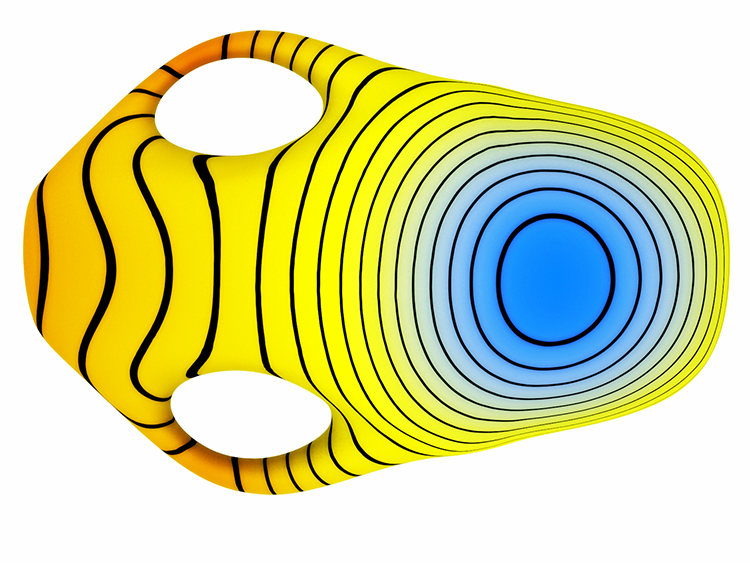}
        \hspace{5pt}
    }
	\caption{The (a) shortest-path and (b) commute-time distance measured from a source point on a $2$-dimensional surface embedded in $\Rspace^3$. Blue indicates the regions closest to the source.}  
	\label{fig:metrics}
\end{figure}

These distance metrics are illustrated in Figure~\ref{fig:metrics}. This illustration shows the distance from a point source to all other locations on the surface. From this we see commute-time distance produces a smoother gradient than shortest-path distance.

\vspace{-2pt}

%% file: sec-methods-topo.tex
\subsection{Extracting Topological Features}
\label{sec.approach.topology}

To extract topological features from each graph instance, we apply persistent homology to its metric space representation. To describe our process, we first briefly review persistent homology. We then describe persistence diagrams, which encode topological features of a given graph instance.  For more details and background on persistence homology see~\cite{EdelsbrunnerHarer2008} and the references within.

\paragraph{Topological features.}
Homology deals with topological features of a space.  
Given a topological space $\Xspace$, 
the $0$-, $1$- and $2$-dimensional homology groups, denoted as 
$\Hgroup_0(\Xspace)$, 
$\Hgroup_1(\Xspace)$ and $\Hgroup_2(\Xspace)$ respectively, correspond to 
(connected) components, tunnels and~voids of $\Xspace$. 

In our context, we care about the $0$- and $1$-dimensional topological features of a graph instance $G_i$, which correspond to $\Hgroup_0$ and $\Hgroup_1$ of its metric space representation. 
These $0$- and $1$-dimensional topological features, roughly speaking, capture \emph{connected components} and \emph{tunnels} formed by vertices in the instances. 

\paragraph{Persistent homology.}
In practice, there might not exist a unique scale that captures topological structures of the data. Instead, we adapt a multi-scale notion of homology, called \emph{persistent homology}, a main tool in topological data analysis, 
to describe the topological features of a space at different spatial resolutions.

Persistent homology typically starts with a finite set of points in a metric space. 
In our setting, each graph instance $G_i$ is associated with a metric space, where vertices in $G_i$ form a finite set of points $S$, and $d_i$ encodes the pairwise distance among points in $S$.

\begin{figure}[!bht]
	\centering
	\includegraphics[width=0.7\linewidth]{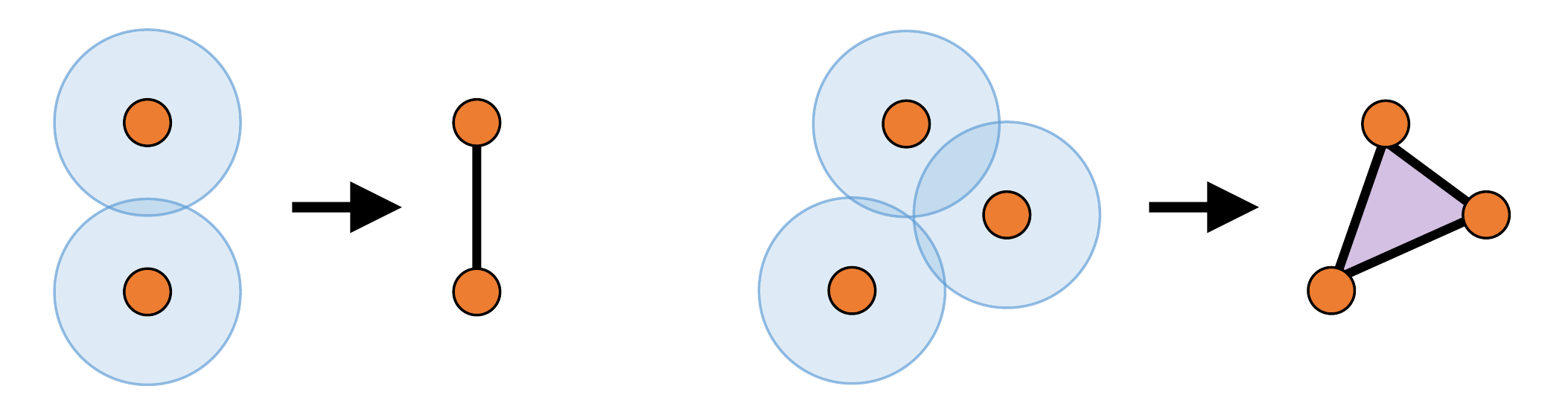}
	\caption{Edges (left) and triangles (right) in a Rips complex.}
	\label{fig:VR-conditions}
\end{figure}

We then apply a geometric construction, such as a Rips complex, on the point set $S$, that describe the combinatorial structure among the points. 
For a real number $r > 0$, a \emph{Rips complex}, denotes as $\Rips(r)$, is formed by considering a set of balls of radius $r/2$ centered at points in $S$.  a $1$-simplex (an edge) is formed between two points in $S$ if and only if their balls intersect (see Figure~\ref{fig:VR-conditions} left). A $2$-simplex (a triangular face) is formed among three points if the balls intersect between every pair of points (see Figure~\ref{fig:VR-conditions} right). 

Given a finite point set $S$ from $G_i$, continuously increase the diameter forms a $1$-parameter family of nested unions of balls; and correspondingly we obtain a $1$-parameter family of nested Rips complexes, referred to as a \emph{Rips filtration}. 
Let $0 = r_0 \leq r_1 \leq r_2 \leq \cdots \leq r_m$ denote a finite sequence of increasing diameter. 
The Rips filtration $\filtration_i$ (of $G_i$) is a sequence of Rips complexes connected by inclusions,   
$\Rips(r_0) \to \Rips(r_1) \to \Rips(r_2) \to \cdots \to \Rips(r_m).$

\begin{figure}[!tb]
\begin{center}
\includegraphics[width=0.91\linewidth]{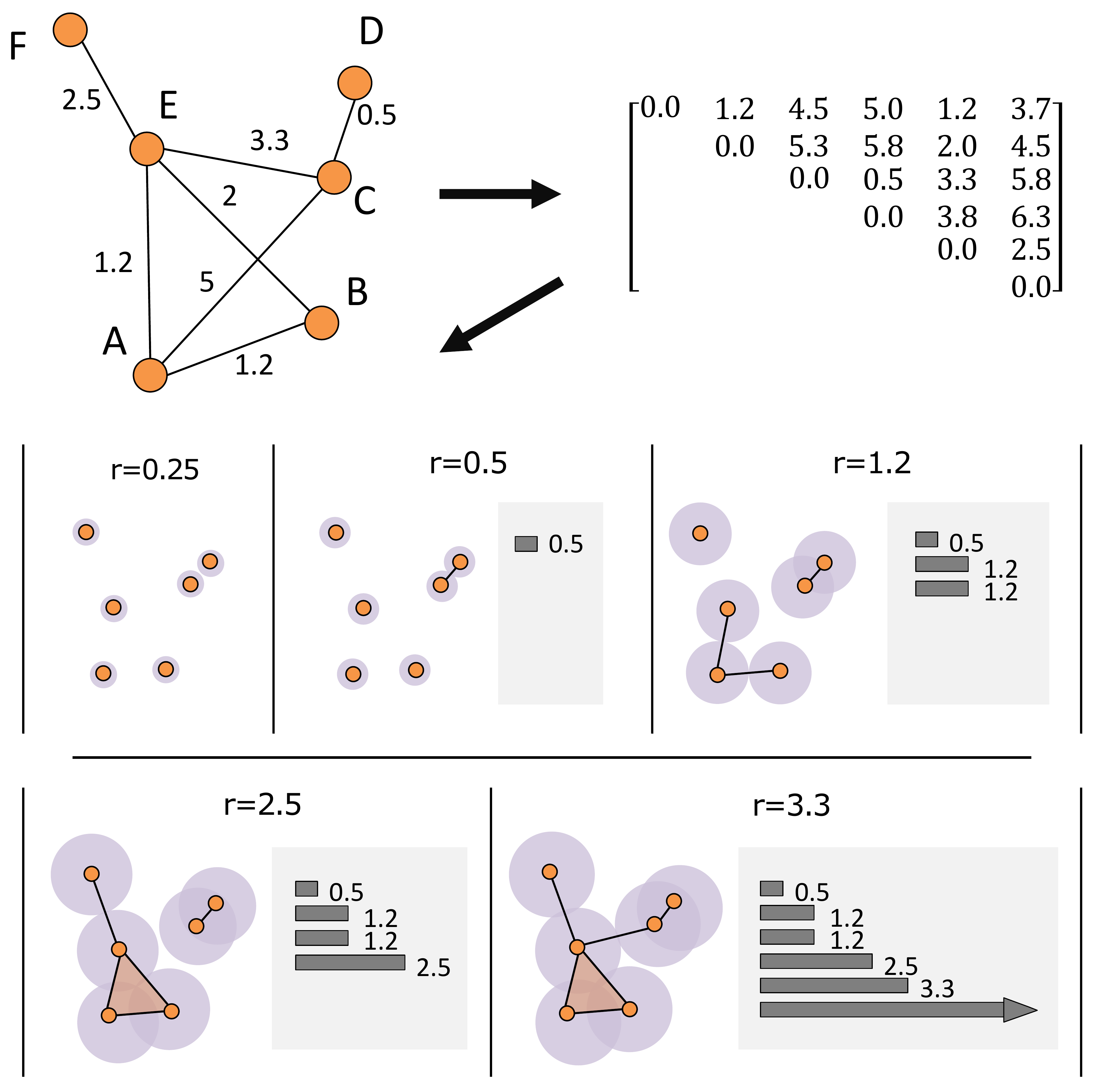}
\end{center}
\caption{Constructing a Rips filtration from a distance matrix on a graph. The numbers above each Rips complex indicates the diameter at which the complex is computed. The corresponding 0-persistence diagrams are shown in the gray box to the right of each complex.}
\label{fig:filtration_example}
\end{figure}

Figure~\ref{fig:filtration_example} shows a Rips filtration defined on an example graph equipped with a shortest-path metric.

Applying homology to a Rips filtration, the homology groups are connected from left to right by homomorphisms induced by inclusions, $\Hgroup(\Rips(r_0)) \to \Hgroup(\Rips(r_1)) \to \Hgroup(\Rips(r_2)) \to \cdots \to \Hgroup(\Rips(r_m)).$

Topological features appear and disappear as the diameter increases: 
when a topological feature appears, that is, a cluster or a tunnel forms, this is called a \emph{birth} event; 
when a topological feature disappears, that is, two clusters merge into one or a tunnel is filled, it is called a \emph{death} event. 
The \emph{persistence} of a topological feature, is the time difference between the death and the birth event.

\paragraph{Persistence Diagrams.}
Topological features of a graph instance and their persistence are recorded by pairing their birth and the death events as a multi-set of points in the plane, called the \emph{persistence diagram} (see~\cite{EdelsbrunnerHarer2010}).

Each topological features is represented as a point $(u, v)$, where $u$ is the \textit{birth time}, and $v$ is the \textit{death time} of the feature. Certain feature may ``live'' forever, in that case, they are assigned a death time of $\infty$. 
Therefore, persistence diagram contains a multi-set of points in the extended plane (i.e.~$(\Rspace \cup \pm\infty)^2$). 
For technical reasons, we add the points on the diagonal to the diagram, each with infinite multiplicity. 
The \emph{persistence} of the pair $(u, v)$ is simply $|v-u|$. Features with higher persistence carry more significant topological information. Features with low persistence are typically considered to be noise.
A persistence diagram can be visualized as persistence barcodes~\cite{Ghrist2008} (see Figures~\ref{fig.pipeline} and \ref{fig:filtration_example}), where each \emph{bar} starts at time $u$ and ends at time $v$. 
We are interested in $0$- and $1$-dimensional topological features, so we consider the $0$- and $1$- persistence diagrams, denoted as $\PD_0(\filtration_i)$ and $\PD_1(\filtration_i)$, respectively.

%% file: sec-methods-pd.tex
\subsection{Comparing Sets of Topological Features}
\label{sec.approach.compare}

A persistence diagram can be thought of as a summary of topological features of a graph instance $G_i$. 
To quantify the structural difference between two instances $G_i$ and $G_j$, we compute the bottleneck and Wasserstein distances between their persistence diagrams.

Given two persistence diagrams $X$ and $Y$, let $\eta$ be a bijection between points in the diagram. 
The bottleneck distance~\cite{EdelsbrunnerHarer2010} is defined as, 
\begin{equation}
W_{\infty}(X,Y) = \inf_{\eta: X \rightarrow Y} \sup_{x \in X} \left\lVert x-\eta(x) \right\rVert_\infty.
\end{equation}
The Wasserstein distance is,
\begin{equation}
W_q(X,Y) = \displaystyle \left[ \inf_{\eta:X \rightarrow Y}  \Sigma_{x\in X} \left\lVert x-\eta(x) \right\rVert^q_\infty \right]^{1/q}, 
\end{equation}
for any positive real number $q$; in our setting, $q=2$.

The set of points in the persistence diagram can be considered as a \emph{feature vector}, where the \emph{feature space} consists of all persistence diagrams for the time-varying graph $G$. 
Given all pairwise distances between persistence diagrams, classical multidimensional scaling (MDS) is then used to reduce the dimensionality of the feature vectors for visualization, and to identify the instances where topologically interesting events occur.

%% file: sec-methods-vis.tex
\subsection{Visualization}
\label{sec.approach.vis}

The design goal of our interactive visualization tool is to provide insights about variation in the structural properties of time-varying graphs. In this way, we hope to identify time periods of uniform behavior (low variation) and outlier behavior (instances of high variation). Our visualization tool provides a number of capabilities to support this form of investigation.

\paragraph{Timeline.} The timeline view uses the horizontal axis to represent time and the vertical axis to represent the first dimension returned by applying classical MDS to the space of persistence diagrams.  
This in essence highlights the dissimilarity between graph instances.
Each point on the timeline represents a single instance of the time-varying graph. 
The points are colored using cyclic colormaps, such as the time-of-day colormap of Figure~\ref{fig.teaser} or the day-of-the-week colormap of Figure~\ref{fig.email.outliers}.

\begin{figure*}[!t]
	\centering
	\includegraphics[width=0.9\linewidth]{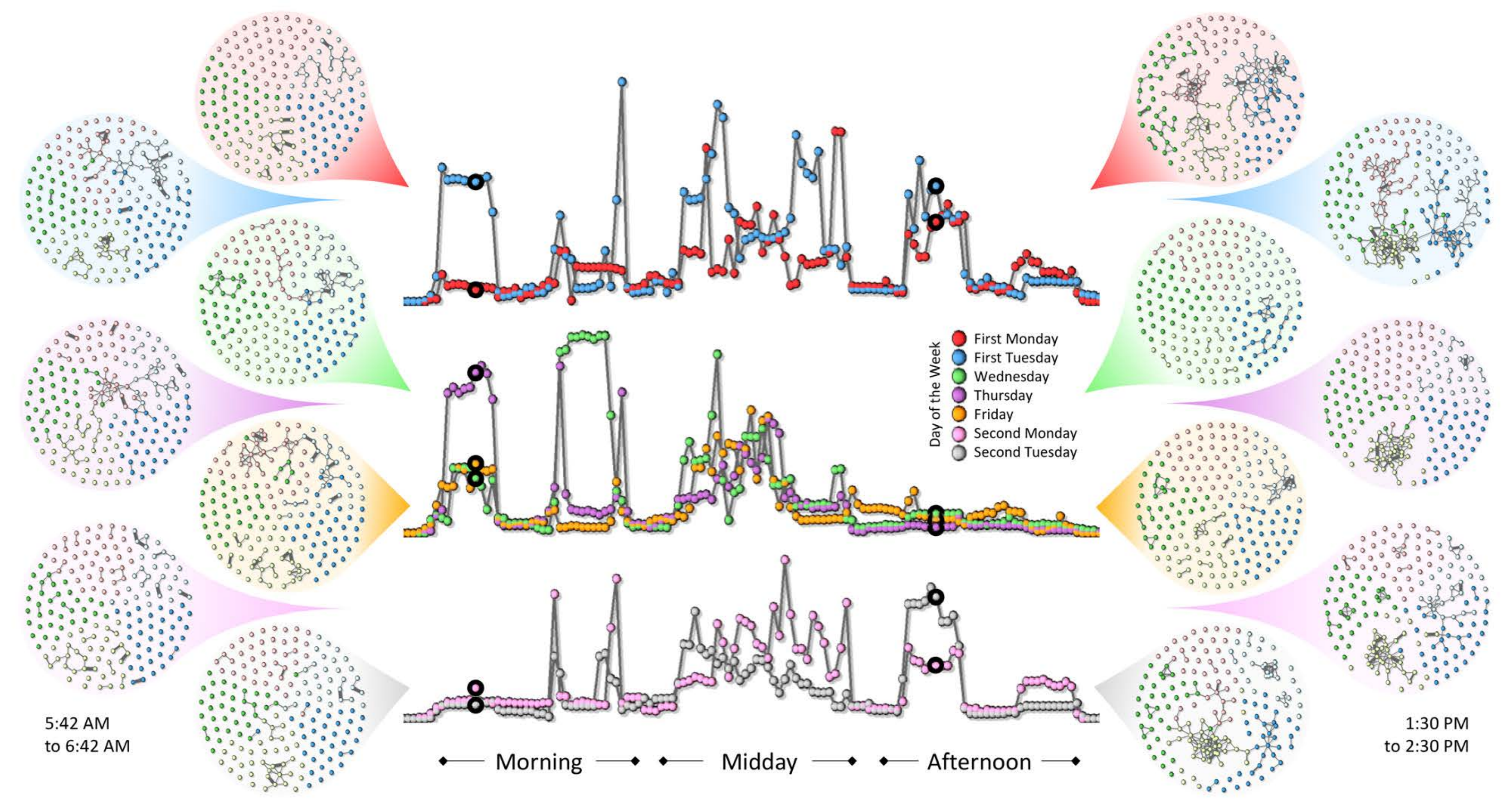}
\caption{Timeline comparison for the 7 weekdays of the High School Communications dataset. The timeline was generated by comparing the commute-time 0-dimensional homological features of the time-varying network using Wasserstein distance. Additionally, two sets of graphs, each from the same time for 7 different days are provided. These graphs validate the different levels of communication visible using our approach.}
\label{fig.hs.time_of_day_compare}
\end{figure*}

\paragraph*{Cyclic Patterns.} Two techniques are available for showing repetitive patterns in the data, both being variations of the timeline. The first technique simply splits the data based upon a user-specified period length. Each period is colored uniquely. Figure~\ref{fig.hs.time_of_day_compare} shows an example of this. For the second technique, the time periods are clustered based upon their $\ell^2$-norm using $k$-means clustering with a user specified $k$. Figure~\ref{fig.email.clusters} shows an example of this where the points are colored by day-of-the-week. 

\begin{figure}[!b]
 \begin{center}
  \includegraphics[width=0.91\linewidth]{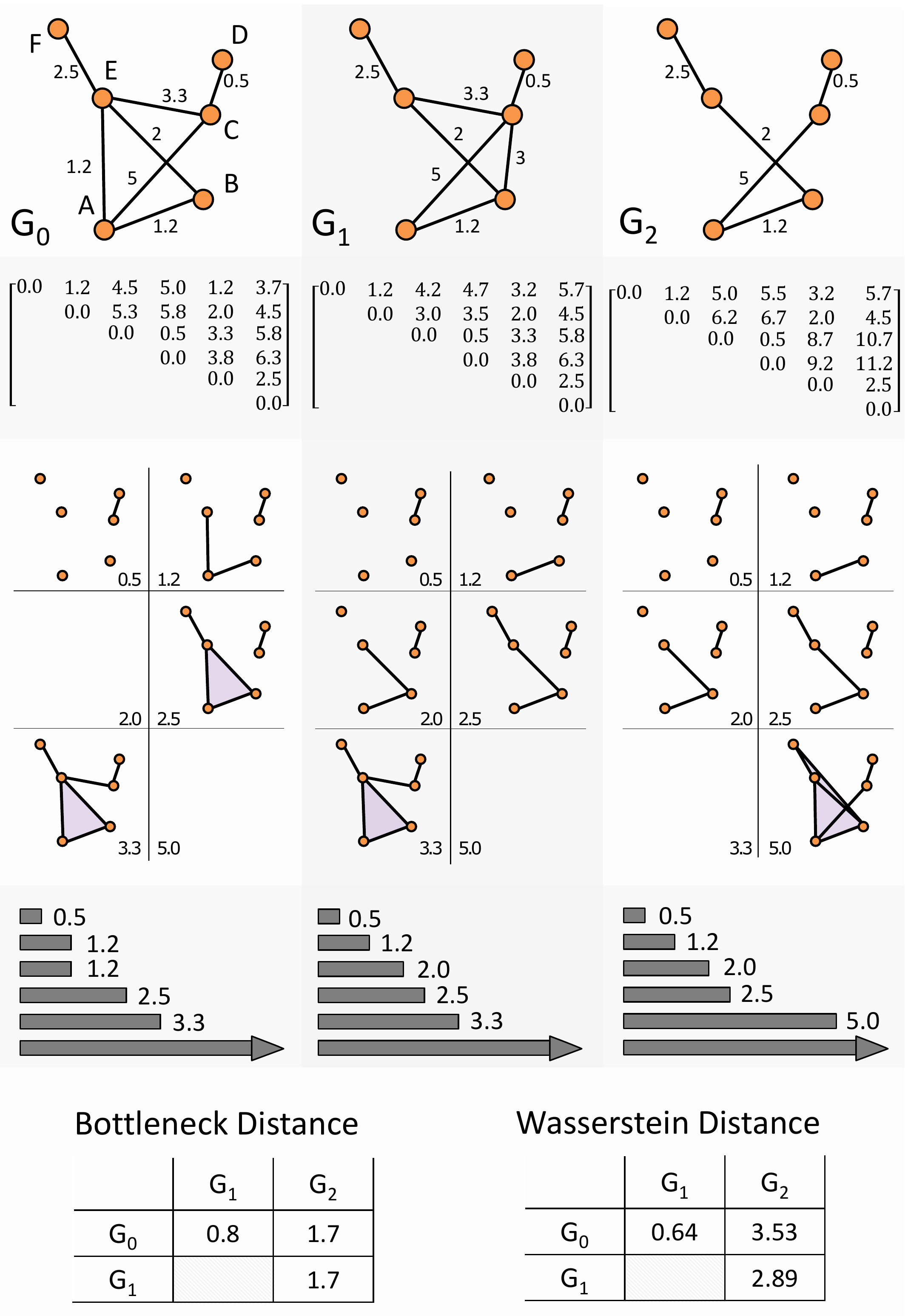}
 \end{center}
\caption{From left to right, 1st row: three weighted graph instances $G_0$, $G_1$ and $G_2$ representing a time varying graph. 2nd row: each graph instance is embedded into a metric space, represented by a shortest-path distance matrix. 3rd row shows the filtrations in which topologically significant events occur, resulting in persistence barcodes in the 4th row. 5th row: the persistence diagrams are compared pairwise using bottleneck and Wasserstein distance.}
\label{fig:graph_sequence}
\end{figure}

\paragraph{Graph Visualization.} For investigating the behavior of specific graph instances, the instances are displayed by two visualization mechanisms. The first is a node-link diagram created using a force-directed layout. If categorical information is available (such as in Figure~\ref{fig.teaser}), the nodes are colored by those categories. For 1-dimensional topological features, nodes can be parameterized around the tunnel using a 1-dimensional cyclic parameterization~\cite{SilvaMorozovVejdemo-Johansson2009, WangSummaPascucci2011}, and colored accordingly. An example of this is seen in Figure~\ref{fig.hs.1dim}. In other cases, nodes receive a fixed color. 
The second mechanism visualizes the persistence diagram for a given graph instance using its barcodes (see fourth row of Figure~\ref{fig:graph_sequence}). The barcode is a variation on a bar chart that represents the birth and death of all topological features in the graph.

%% file: sec-methods-example.tex
\subsection{Example}

We provide an illustrative example of our pipeline in Figure~\ref{fig:graph_sequence}.

In step 1 (1st row), a time-varying graph $G$ is given as sequence of graph instances, where each instance is a connected, weighted graph. 
In step 2 (2nd row), each graph instance is embedded in a metric space by calculating a distance matrix using the shortest-path metric. In step 3 (3rd row), the distance matrix is used to compute a series of filtrations. In reality, additional filtrations are created, but we only shows those that produce topological events, in this example only 0-dimensional features. The step 4 (4th row), the $0$-dimensional persistence diagrams of the filtrations are extracted and shown as barcodes. The final step (5th row) consists of computing the distances between these diagrams using bottleneck and Wasserstein distances. 

The bottleneck or Wasserstein distance as \emph{persistence-based similarity measure} helps to quantify topologically similarity between a pair of instances. 
For example, under both distances, $G_0$ and $G_1$ are much closer to one another than either $G_0$ and $G_2$ or $G_1$ and $G_2$.

%% file: sec-evaluation.tex
\section{Case Studies}
\label{sec:evaluation}

To validate our approach, we look at case studies of two publicly available datasets. Both are communication networks, one involves interpersonal communication of high school students, and the other contains e-mail communications between researchers. These case studies help demonstrate how our approach can identify cyclic patterns in data, deviations from patterns, and one-time events in time-varying graphs.

Our pipeline requires a number of tools for processing. Graph processing and metric space embedding are coded using Python. Persistent homology calculations and the bottleneck and Wasserstein distances are computed using Dionysus\footnote{\url{http://www.mrzv.org/software/dionysus/}}. 
Finally, visualizations are implemented using Processing\footnote{\url{https://processing.org/}}.

\subsection{High School Communications}

The High School Communications dataset~\cite{fournet2014contact} is a time-varying graph that tracks the contact between high school students. The data was collected for 180 students in 5 classes over 7 school days in November 2012 in Marseilles, France. The graph tracks Monday through Friday of the first week and Monday and Tuesday of the following week.

We compute both shortest-path and commute-time distances and both $0$- and $1$-dimensional persistence diagrams. Then, both the bottleneck and Wasserstein distances are used to compare persistence diagrams. We present a small set of configurations and draw a few conclusions from them. Many similar conclusions have been identified in other configurations that are not shown.

\begin{figure}[!b]
  \centering
  \includegraphics[width=0.84\linewidth]{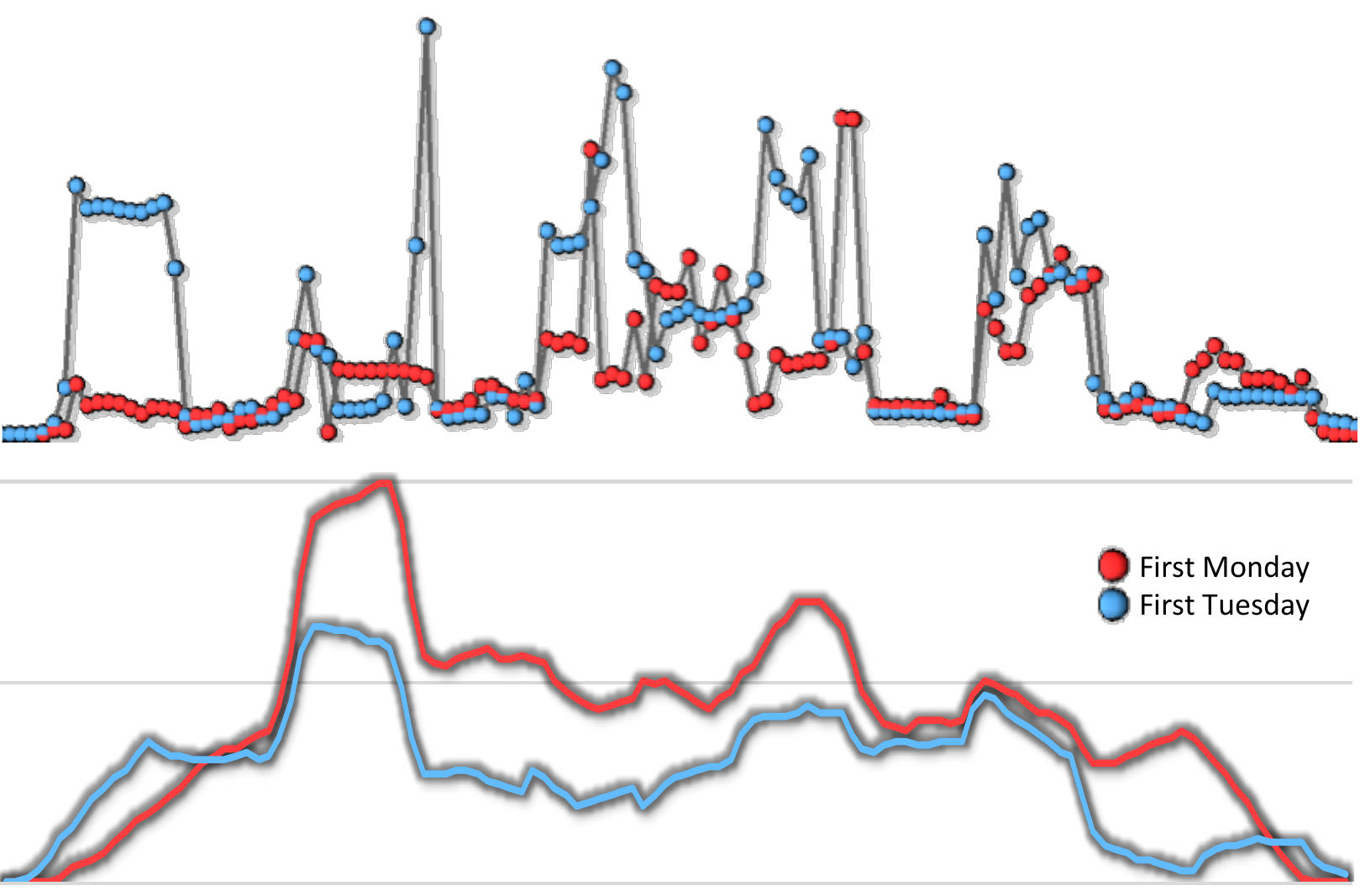}
\caption{Top: Persistent homology timeline for the first Monday and Tuesday of the High School Communications dataset. Bottom: Timeline counting the number of events (sum of all weights) in each graph instance. The timeline shows how different features can be identified in our approach as compared to edge counts alone.}
\label{fig.hs.edge_day}
\end{figure}

\subsubsection{An Average Day} 
First, to examine an average day of communication, we look at the $0$-dimensional features of the first Monday of the dataset in Figure~\ref{fig.teaser}. In this figure, commute-time is used to generate persistence diagrams and bottleneck distance is used to compare diagrams. In this figure, a number of phases can be seen. In the early and late hours, no interactions occur (e.g., time A). As the school day begins at time B, light, loosely-connected communications begin. By mid-morning (time C), class MP*1, PC, PC*, and PSI* are all interacting heavily within and between groups. Midday (times D \& E), shows classes heavily interacting once again. Early afternoon (time F) shows mostly within communications for classes PC, PC*, and PSI* and within and between communications for MPI*1 and MPI*2. Finally, the end of the day, time G, shows much sparser group communications.

\begin{figure}[h]
 \centering
  \includegraphics[width=0.89\linewidth]{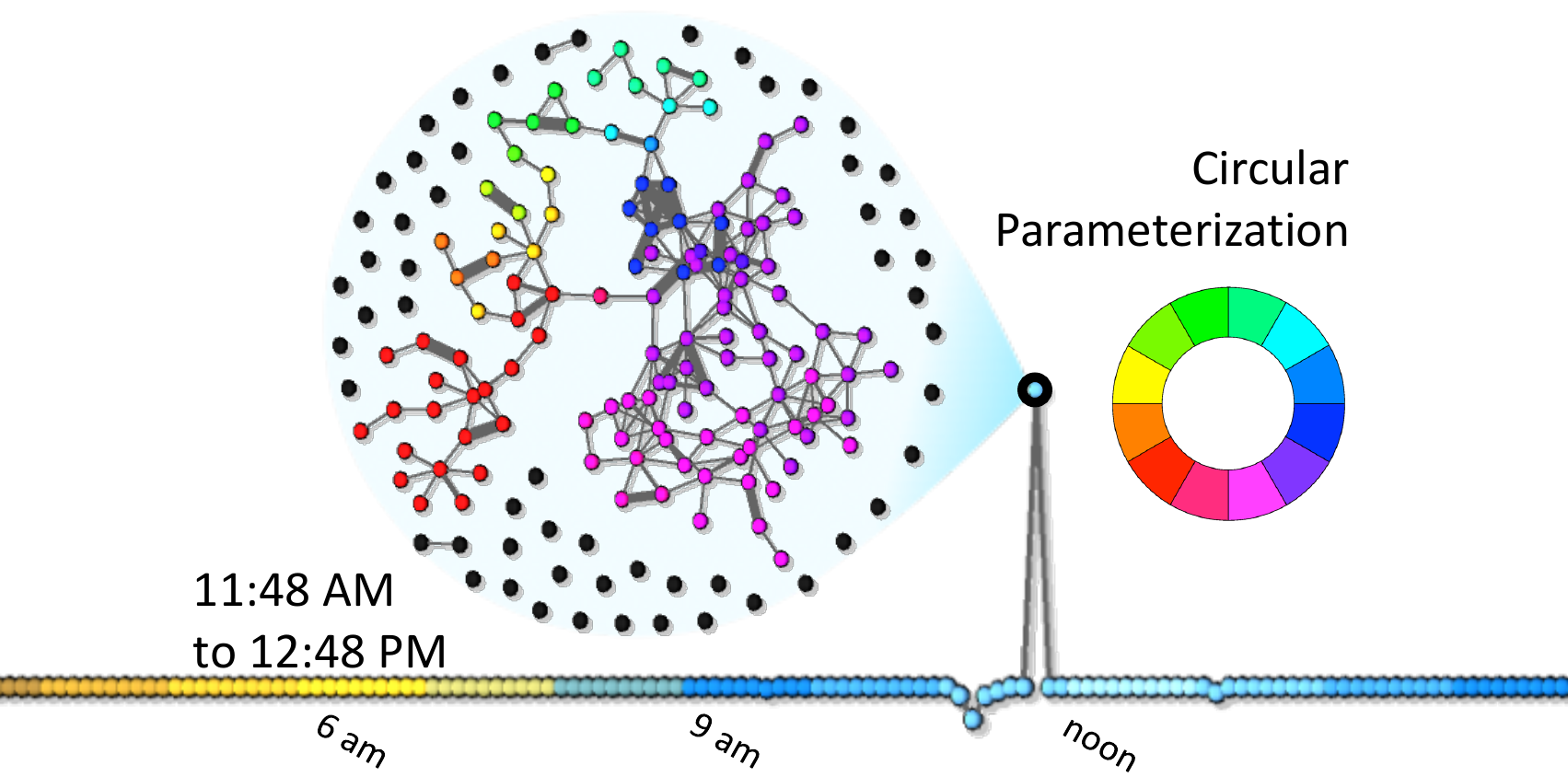}
\caption{Timeline of the High School Communications dataset for 1-dimensional features. The timeline was generated by comparing the commute-time features using bottleneck distance. The single outlier is a graph with a high persistence cycle. To highlight that feature, the graph is parameterized and visualized with a cyclic rainbow colormap~\cite{WangSummaPascucci2011}.}
\label{fig.hs.1dim}
\end{figure}

\begin{figure}[!b]
 \centering
 	\begin{minipage}{.04\linewidth}
 		\vspace{0pt}\subfigure[\label{fig.email.bottle_v_wasser.b0}]{}
    
    	\vspace{0pt}\subfigure[\label{fig.email.bottle_v_wasser.b1}]{} 
    
    	\vspace{2pt}\subfigure[\label{fig.email.bottle_v_wasser.w0}]{} 
    
    	\vspace{4pt}\subfigure[\label{fig.email.bottle_v_wasser.w1}]{}
    \end{minipage}
 	\begin{minipage}{.93\linewidth}
	  \includegraphics[width=0.97\linewidth]{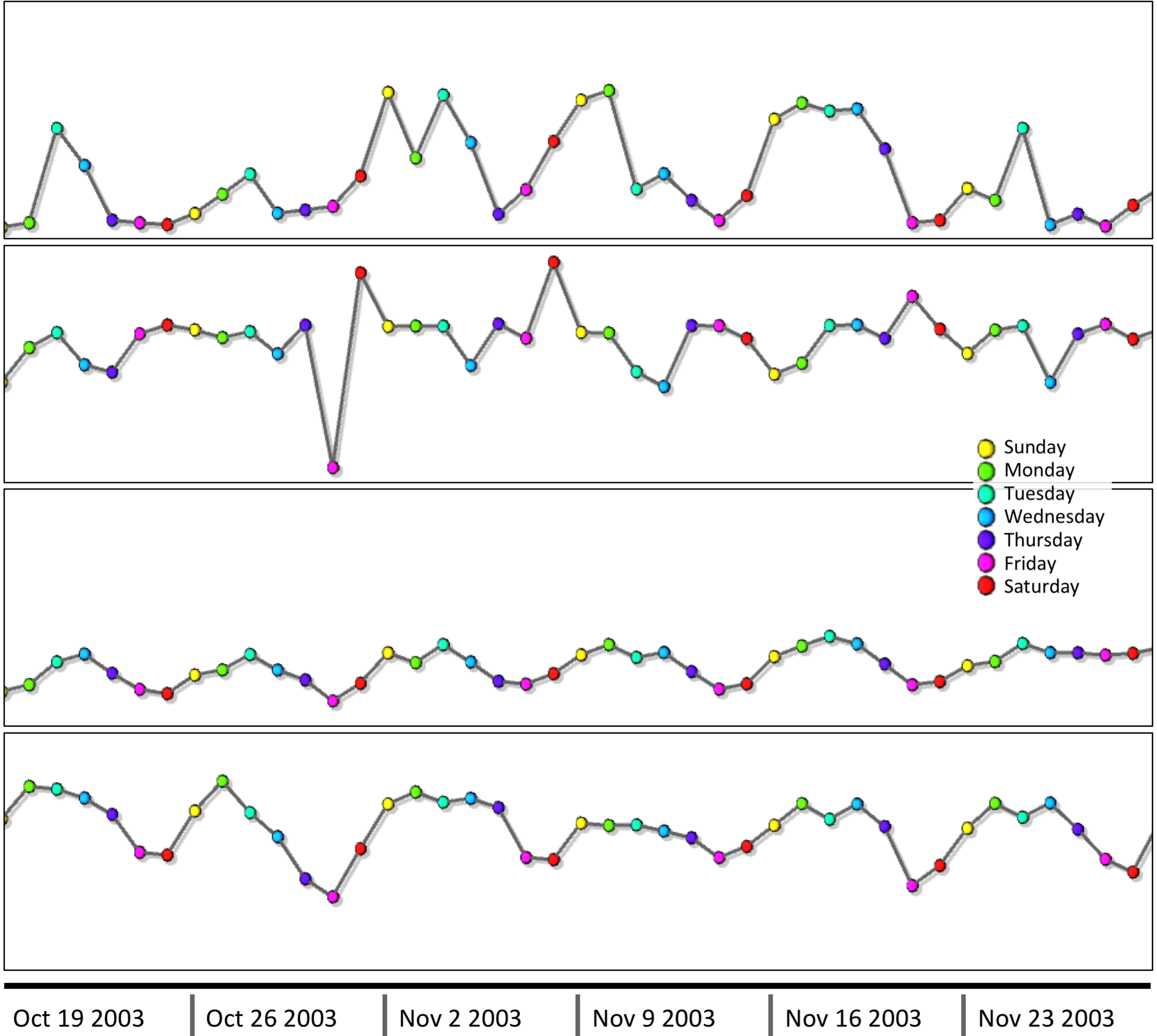}
    \end{minipage}
\caption{Comparing shortest-path bottleneck ((a) and (b)) and Wasserstein ((c) and (d)) distance on 0-dimensional ((a) and (c)) and 1-dimensional ((b) and (d)) features in the EU E-Mail dataset. Since bottleneck distance captures the most perturbed feature, the result may be noisy. Wasserstein distance captures variation across all features in the graph resulting in a smoother pattern.}
\label{fig.email.bottle_v_wasser}
\end{figure}

\begin{figure*}[!t]
  \centering
  \includegraphics[width=0.91\linewidth]{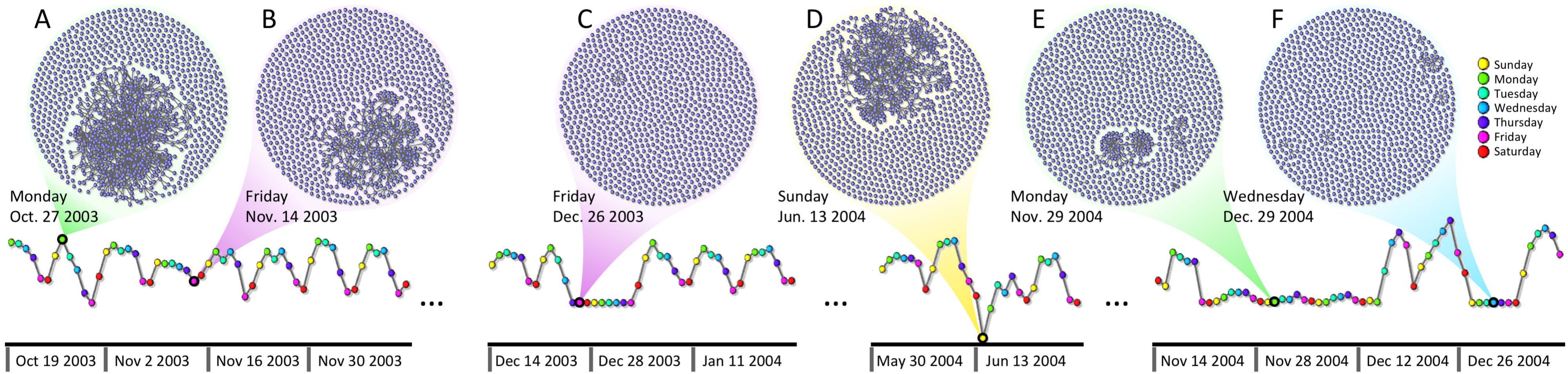}
\caption{Highlights from the EU E-Mail dataset using the shortest-path Wasserstein distance on 1-dimensional persistence diagrams. A \& B show graphs from a timeframe of normal weekly cyclic activity. C \& F show timeframes of limited activity from December of 2003 and 2004 during the Christmas and New Years Holidays. D shows an unexpected boost in activity on June 13, 2004 that is correlated with the release of results for the EU Parliamentary Election. E shows a 3-4 week period of low activity in November and December of 2004. We could not identify any externally correlated event to explain this occurrence.}
\label{fig.email.outliers}
\end{figure*}

\subsubsection{Comparison with Other Days}
While observing patterns within a single day is interesting, comparing Monday with other days can help to better identify regular and irregular daily behavior. Figure~\ref{fig.hs.time_of_day_compare} shows just such a comparison; it uses commute-time to generate $0$-dimensional persistence diagrams, and Wasserstein distance to compare diagrams. 

The top chart of Figure~\ref{fig.hs.time_of_day_compare} compares the first Monday and the first Tuesday. Ignoring outlier graph instances, two main differences can be observed. First, the early morning of Tuesday shows different levels of activity than Monday. This can be confirmed by looking at examples from those days. Figure~\ref{fig.hs.time_of_day_compare} (top left) shows example graphs from Monday and Tuesday morning. Secondly, at both the beginning and end of midday, Tuesday shows higher activity than Monday. 

The middle chart of Figure~\ref{fig.hs.time_of_day_compare} compares Wednesday, Thursday, and Friday. In this chart, Wednesday and Friday show more early morning activity than Monday, but Thursday shows activity levels similar to Tuesday. Individual graph instances of the time-varying graph from this timeframe can be seen in Figure~\ref{fig.hs.time_of_day_compare} (middle left). Late morning shows that Wednesday is extremely active, while Thursday and Friday are mostly inactive. Midday across all three days remains similar. Finally, the afternoons of all three days are similarly inactive. Sample graphs for this timeframe can be seen in Figure~\ref{fig.hs.time_of_day_compare} (middle right).

The bottom of Figure~\ref{fig.hs.time_of_day_compare} shows the second Monday and Tuesday. These days show almost no morning activity (also see Figure~\ref{fig.hs.time_of_day_compare} (bottom left)) and normal midday activity. Early afternoon shows midrange and high activity for Monday and Tuesday, respectively. Graphs associated with these activity levels can be seen in Figure~\ref{fig.hs.time_of_day_compare} (bottom right). 

As a means to compare results to a more traditional analytic, Figure~\ref{fig.hs.edge_day} bottom is a timeline that captures the number of interaction events for a given graph instance in the time-varying graph (i.e.\ the sum of the weights). Comparing this chart to that in Figure~\ref{fig.hs.edge_day} bottom, it is clear that our approach captures a \textit{different} type of behavior than edge counting alone.

\subsubsection{1-Dimensional Topological Features}

The High School Communications dataset ultimately contains very few 1-dimensional topological features, the majority of which have low persistence. The one-time exception, which appears on the first Monday, can be seen in Figure~\ref{fig.hs.1dim}. Between 11:48 am and 12:48 pm a high persistence $1$-dimensional pattern appears in the graph. The nodes of the graph are parameterized using that feature and visualized using a cyclic rainbow colormap. The graph shows a large tunnel (loop) towards the upper left.

\subsection{EU Research Institution E-Mail}

The EU Research Institution E-mail~\cite{leskovec2007graph}\footnote{\url{http://snap.stanford.edu/data/email-Eu-core.html}} dataset is an anonymized time-varying graph tracking e-mails between members of ``a large European research institution''. We have used the smaller of the available networks containing 986 nodes and 332,334 temporal edges. The graph tracks the activity for 803 days. A period of about 200 days is missing towards the end of the dataset, so we have analyzed the first 500 days. A single graph instance is created per day and shared $45\%$ overlap with neighboring days. Once again, edge weight is chosen by counting the number of communications during the graph instance.

\subsubsection{Bottleneck vs.\ Wasserstein Distance}

The bottleneck and Wasserstein distance both capture important but distinct differences among sets of topological features. 
Intuitively, the bottleneck distance ($p = \infty$) captures the most perturbed topological feature (or the extreme behavior); while the Wasserstein distance ($p = 2$) captures the perturbation across all features (or the average behavior). Figure~\ref{fig.email.bottle_v_wasser} shows how this impact the analysis of the EU E-Mail dataset. For $0$-dimensional (Figure~\ref{fig.email.bottle_v_wasser.b0}) and $1$-dimensional (Figure~\ref{fig.email.bottle_v_wasser.b1}) bottleneck distances, the result is noisy, as the value captured has the most variation. For $0$-dimensional (Figure~\ref{fig.email.bottle_v_wasser.w0}) and $1$-dimensional (Figure~\ref{fig.email.bottle_v_wasser.w1}) Wasserstein distances, the result is smoother, since it encodes the perturbations across all features. For our analysis of the EU E-mail data, this property is more desirable.

\subsubsection{Revealing Cyclic Patterns}

Upon investigating the data, cyclic patterns were immediately apparent with all configurations of the Wasserstein distance ($0$- \& $1$-dimensional features and shortest-path \& commute-time).  Figure~\ref{fig.email.outliers} A \& B show the 1-dimensional shortest-path version, where the cyclic patterns are most prominent (also see supplemental material for the complete 1-dimensional feature timeline). 

\begin{figure}[h]
 \centering
  \includegraphics[width=0.93\linewidth]{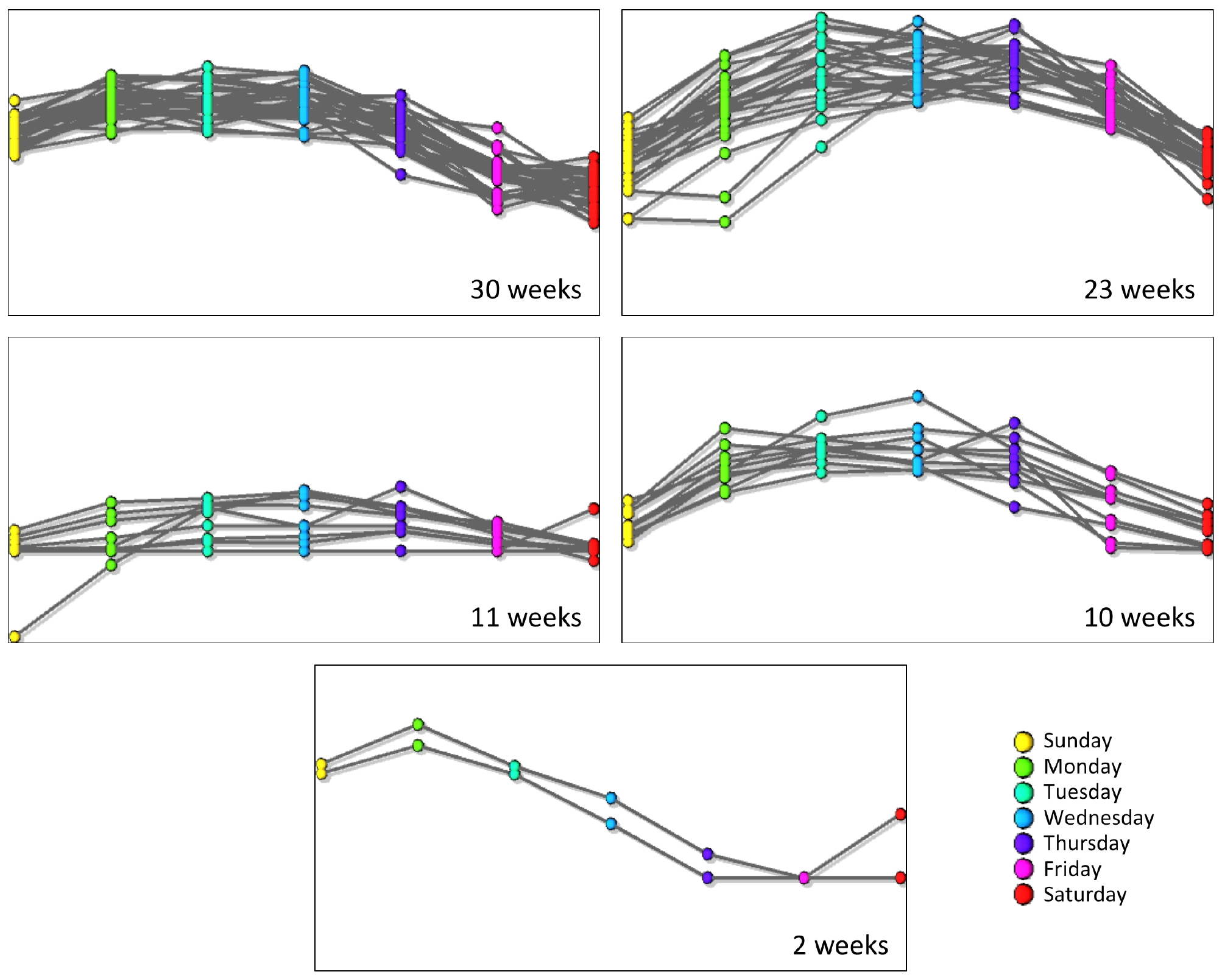}
\caption{Clustering of the weekly behavior in the EU E-Mail dataset using the shortest-path Wasserstein distance on 1-dimensional features. The clusters shows 4 primary patterns and 1 outlier pattern (bottom). The number of weeks in each cluster is listed in the lower right.}
\label{fig.email.clusters}
\end{figure}

It is notable that this pattern is related to the natural cycle of the week. To identify the pattern of the ``standard'' week, we divided the data into 7 day segments and used k-means clustering to group similar weeks. Figure~\ref{fig.email.clusters} shows the result with 5 clusters. Each of the 5 clusters shows a version of the typical week for this institution.

\subsubsection{One-time Events}

When looking at the entire timeline (see supplemental material), a number of one-time events are easily discovered. Figure~\ref{fig.email.outliers} C \& F are two such events. During these time periods very little activity is present in the graphs. These times happen to be the last week of December and first few days of January, during the Christmas and New Year's holidays. Figure~\ref{fig.email.outliers} D is a one time event that shows an extreme increase in activity for a 1-2 day period. After entering the date, June 13, 2004, into Google, we discovered that this day corresponds to the release day of the results for the EU Parliamentary Election. Finally, Figure~\ref{fig.email.outliers} E shows a 3-4 week period of significantly decreased activity. Despite our best efforts, we could not identify a major external event that would have caused such a reduction, and since the data is anonymized, we could not identify the institution to investigate a local or internal cause.

%% file: sec-discussion-intuitiveness.tex
\section{Discussion}
\label{sec:discussion}

In the previous section, we construct a similarity measure between two graph instances of a time-varying graph by utilizing the bottleneck or 
Wasserstein distance between their persistence diagrams, which encode the topological features associated with each instance. 

However, one might ask: why persistent homology? We argue that using topological data analysis and in particular, persistent homology, to study graphs, have complementary benefits and offer new insights. In this section, we conduct several experiments to justify our approach. 
In addition, we describe some intuition behind the information encoded by the persistence diagram of a graph, and the distances functions defined on them.

\subsection{Persistent Diagram As a Graph Fingerprint}
\label{Intuitiveness }

Conventional graph-theoretical approaches typical utilize the statistical properties of the vertices and edges, for instance, degrees, connectivity, path lengths to describe the short range and pairwise interactions in the system. 
On the other hand, topological summaries, such as the persistence diagrams, are compressed feature representation of the underlying data, that can capture long range and higher-order interactions.

We test our persistence-based similarity measure against a set of desirable properties for a similarity measure on a graph (the first four conditions are introduced in~\cite{koutra2013deltacon}):
\begin{enumerate}\denselist
\item \textbf{Edge importance:} An edge whose insertion or deletion changes the number of connected components is more important than an edge that does not.
\item \textbf{Weight awareness:} In weighted graphs, the bigger the weight of the removed edge is, the greater the impact on the similarity measure should be. 
\item \textbf{Edge-submodularity:} Changing an edge in a dense graph is less important than changing an edge in an equally-sized, sparse  graph. 
\item \textbf{Focus awareness:}  Random changes in graphs are less important than targeted changes of the same extent.
\item \textbf{Node awareness:} We add an extra condition in this paper, i.e.,~deleting a large number of nodes in a graph has a larger impact than deleting a small number of nodes from the same graph.
\end{enumerate}
We conduct several experiments on synthetic and real-world datasets to test the above conditions.

\begin{figure}[!thb]
  \includegraphics[width=0.98\linewidth]{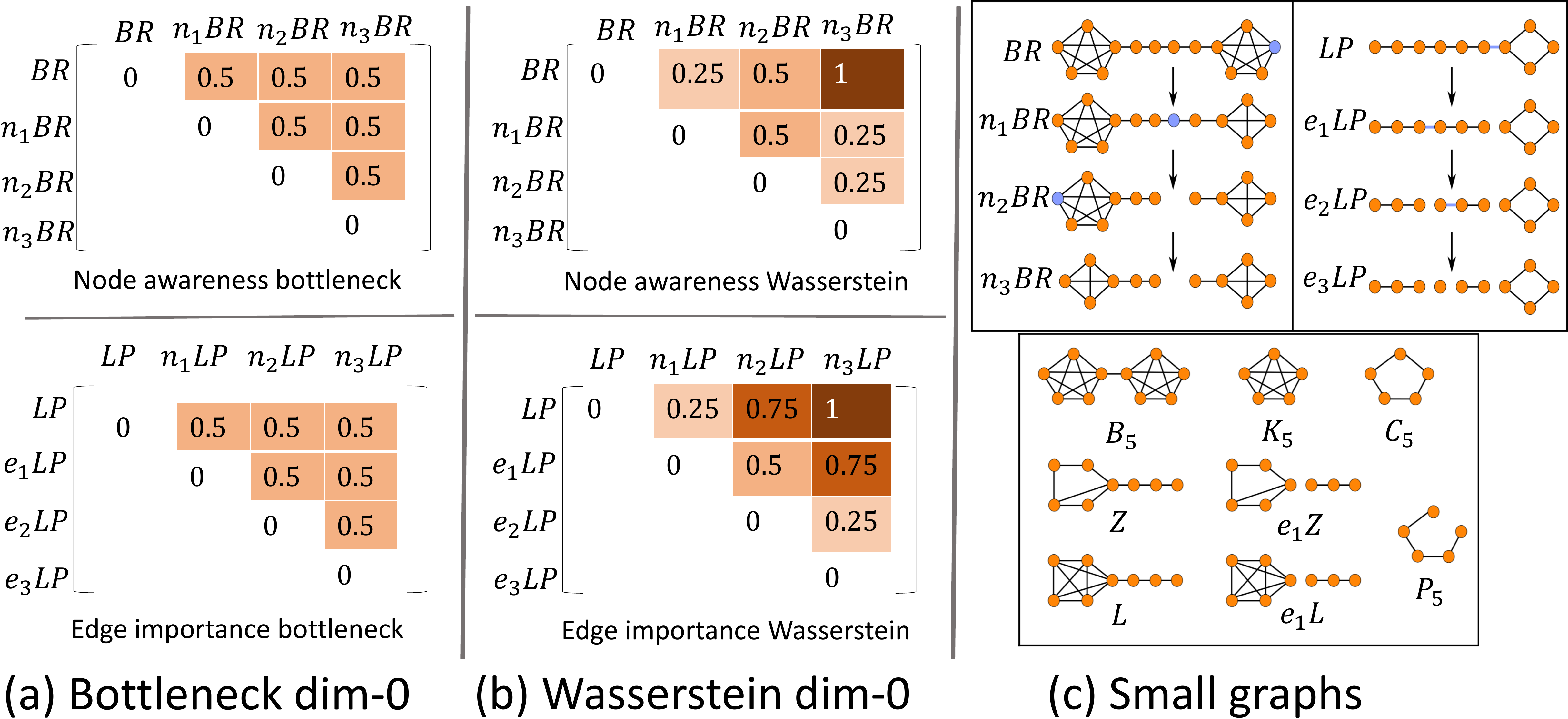}
\caption{Given synthetic, small exemplar graphs in (c), we study the node awareness (property 5, a-b, top) and the edge importance (property 1, a-b, bottom) on these graphs by computing the bottleneck (a) and Wasserstein distances (b) matrix between $PD_0$ of the corresponding graphs. All edge weights are assumed to be $1$.}
\label{fig:matrices}
\end{figure}

For the node awareness (property 5) we consider the graphs $BR$ shown in   Figure~\ref{fig:matrices} (c) top left.
Each of the graphs $n_iBR$ is obtained from the original graph $BR$ by deleting $i$ number of nodes (in blue). The bottleneck and Wasserstein distance matrices of $PD_0$ between these graphs are shown in the top of Figure~\ref{fig:matrices} (a)-(b). The $PD_1$ distance matrices are omitted since their entries are all zeros.
From the matrices in Figure \ref{fig:matrices} (a)-(b) top, we observe that persistence-based similarity measure is sensitive to node deletion, that is, it satisfies \emph{node awareness}, in particular, the Wasserstein distance is more node aware than the bottleneck distance in these examples. 

\begin{figure}[!b]
  \includegraphics[width=0.86\linewidth]{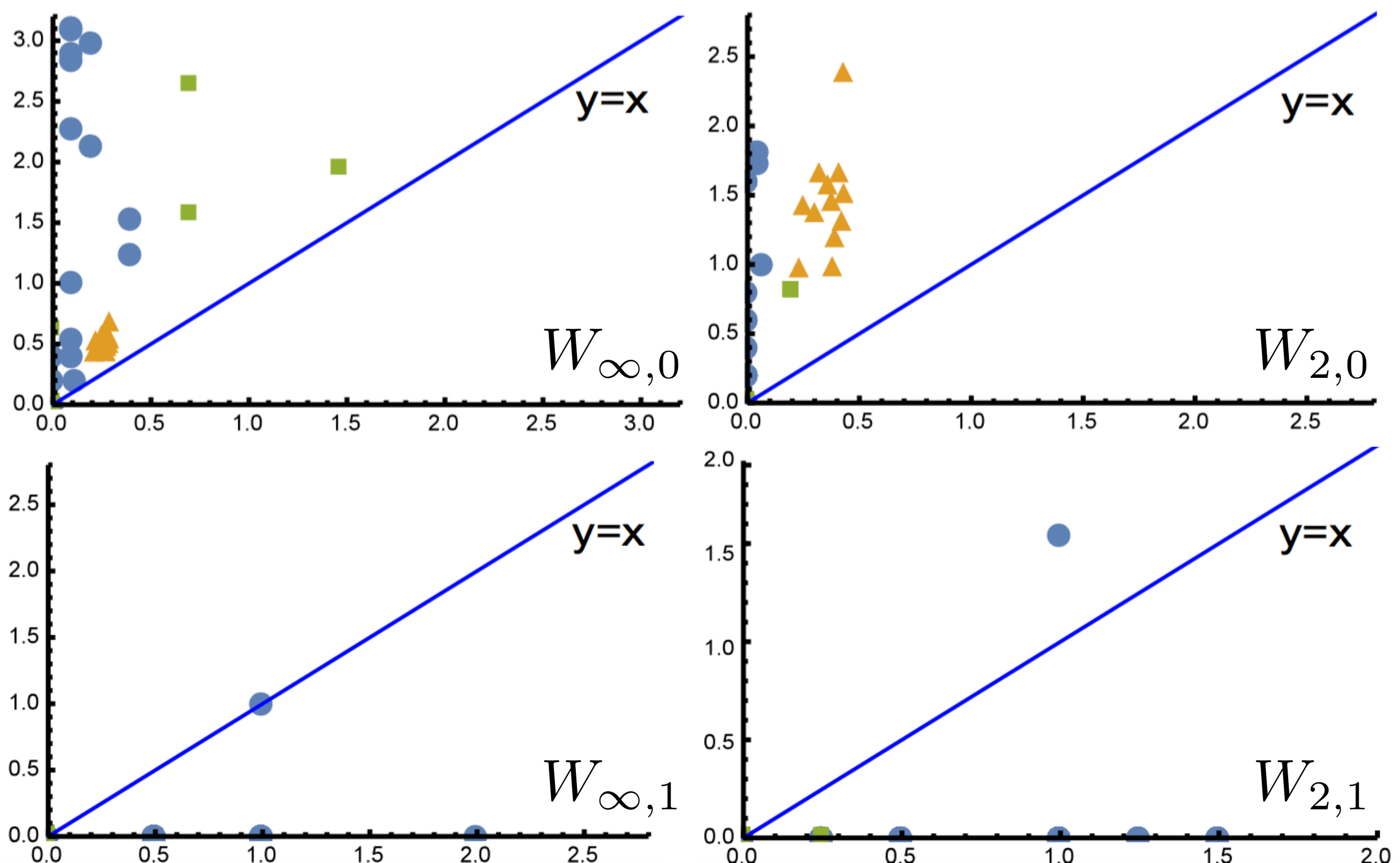}
\caption{Testing  weight awareness (property 2). Points $(W(eA, C^e), W(eA, B^e))$ on and above the diagonal correspond to instances where property 2 is satisfied. 
Three sets of graphs are represented by blue, orange and green points respectively.}
\label{fig:P2}
\end{figure}

Similarly, to test edge importance (property 1) against our similarity measure we delete a set of edges from a graph $LP$, shown in Figure \ref{fig:matrices} (c) top right. The graph $e_iLP$ is obtained from $LP$ by deleting $i$ edges (in blue). The bottleneck and Wasserstein distance matrices of $PD_0$ among these graphs are shown in Figure \ref{fig:matrices} (a)-(b) bottom. 
We observe that our persistence-based similarity measure is sensitive to edge deletions that change the connectivity of the graph, that is, it satisfies \emph{edge importance}. Notice how the the Wasserstein distance is more aware of the level of (dis)connectedness between the graphs.

To test weight awareness (property 2), we run our test on three randomly generated, weighted graphs $A_1 = (V_1, E_1, w_1)$, $A_2 = (V_2, E_2, w_2)$ and $A_3=(V_3, E_3, w_3)$, where $|V_1| = 50$, $|V_2|=60$, $|V_3|=70$, $|E_1| = 200$, $|E_2| = 250$ and $|E_3| = 300$ respectively. 
Each is generated from the $G_{n,m}$ random graph model, where a graph is chosen uniformly at random from the set of all graphs with $n$ nodes and $m$ edges (by setting $n = |V_i|$ and $m = |E_i|$ for $1 \leq i \leq 3$).  
The weights on the edges are drawn uniformly from $(0.1, 1)$. 
For each graph $A_i = (V_i, E_i, w_i)$, we obtain a set of $|E_i|$ modified graphs $B_i^e = (V_i, E_i, u_i)$ by only modifying the weight of an edge $e$ (for all edges) in $A_i$ such that $u_i(e) = w_i(e) + \delta$, where $\delta$ is drawn uniformly randomly from $(4, 5)$; 
similarly, we obtain a set of modified graph $C_i^e = (V_i, E_i, v_i)$  from $A_i$ by only modifying the weight of edge $e$ such that $v_i(e) = w_i(e) + \delta'$, where $\delta'$ is drawn uniformly randomly from $(2, 3)$. 
Let graph $eA_i$ denote the graph obtained from $A_i$ by deleting an edge $e$. Property 2 holds when $W(eA_i,B_i^e)-W(eA_i,C_i^e) \geq 0$ for all $e$ in $A_i$. 

In Figure \ref{fig:P2} we represent the difference $W(eA_i,B^e_i)-W(eA_i,C^e_i)$ by plotting the points $(W(eA_i,C^e_i),W(eA_i,B^e_i))$. Hence, property $2$ holds for a point $(x,y)$ on and above the diagonal (i.e. $y \geq x$). Note that our similarity measure satisfies \emph{weight awareness} for dimension $0$ but violates the condition for dimension $1$. This is due to the fact that $W_{q,1}(eA_i,B^e_i)-W_{q,1}(eA_i,C^e_i)$ for some $e$ captures a creation or a destruction of a cycle.

To test edge-submodularity (property 3), we consider a set of four  graphs $A$, $B$, $C$ and $D$. These graphs share the same number of nodes. Graph $A$ is denser than graph $C$; while graph $B$ and $D$ are obtained from $A$ and $C$ respectively by deleting an edge. 
We test property 3 against four sets of small synthetic graphs in Figure~\ref{fig:matrices} (c) bottom; the results are shown in Table~\ref{table:property3}. 
We see that both Wasserstein and bottleneck on $PD_0$ capture better the changes that occur in a sparser graph than they do on an equally sized denser graph; i.e.~they satisfy \emph{edge-submodularity} in dimension 0.  
However, these distances behave differently on $PD_1$. 
Table \ref{table:property3} shows some negative entries; this is due to the fact that between $C$ and $D$, a cycle is either created or destroyed; while no cycle appears/disappears between $A$ and $B$ (that is, $W(A, B) = 0$). 

\begin{table}[!t]
\centering
\begin{small}
  \centering
\begin{tabular}{|c|c|c|c||c|c|c|c|}
\hline
 \multicolumn{4}{|c||}{Graphs} & %
    \multicolumn{1}{c|}{$W_{2,0}$} &\multicolumn{1}{c|}{$W_{2,1}$} &\multicolumn{1}{c|}{$W_{\infty,0}$} &\multicolumn{1}{c|}{$W_{\infty,1}$} \\ 
\hline
 \multicolumn{1}{|c|}{$A$} &\multicolumn{1}{c|}{$B$} &\multicolumn{1}{c|}{$C$} &\multicolumn{1}{c||}{$D$} & 
    \multicolumn{3}{c}{$\Delta(W)= W(A, B) - W(C, D) $} &\\
\hline
\hline
$C_5$ & $e_1C_5$ &$K_5$ & $e_1K_5$ & 0 & 0.25& 0 & 0.5 \\
\hline
$P_5$ & $e_1P_5$ &$C_5$ & $e_1C_5$ & 0.25 & -0.25& 0.5 & -0.5 \\
\hline
$C_9$ & $e_1C_9$ &$K_9$ & $e_1K_9$ &  0 & 1& 0 & 1 \\
\hline
$P_9$ & $e_1P_9$ &$C_9$ & $e_1C_9$ & 0.25 & -1& 0.5 & -1 \\
\hline
$Z$ & $e_1Z$ &$L$ & $e_1L$ & 0.25 & 0& 0.5 & 0 \\
\hline
\end{tabular}
\end{small}
\hspace{5pt}
  \caption{Testing edge-submodularity (property 3) using the graphs from Figure \ref{fig:matrices} (c).} 
 \label{table:property3}
\end{table}

\begin{figure}[!b]
  \includegraphics[width=0.95\linewidth]{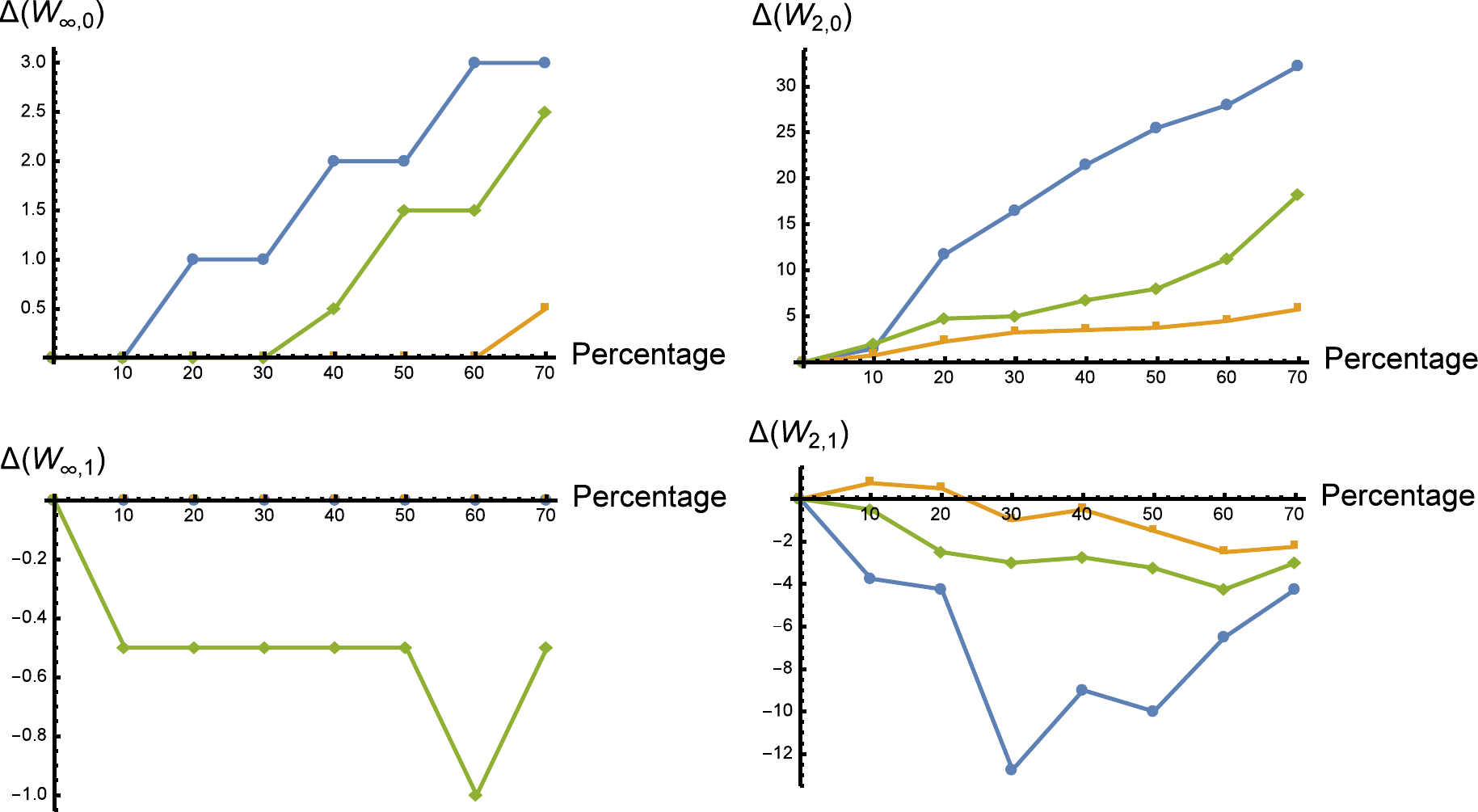}
\caption{Testing focus awareness (property 4). Each colored curve represents a graph among three randomly generated graphs. The difference between the targeted corruption and the random corruption is plotted against the percentage of the deleted edges.}
\label{fig:P4}
\end{figure}

For focus awareness (property 4), we generate three random weighted graphs $A_1$, $A_2$ and $A_3$ following the same $G_{n,m}$ model as before, with $35$, $100$, $120$ vertices, $70$, $500$ and $300$ edges respectively; and all edge weights are chosen uniformly random from $(0.1, 1)$. 
We generate a collection of so-called \emph{corrupted} (i.e.~modified) graphs from the original graph with two types of corruptions: (1) by deleting $10 \%$ to  $70\%$ of random edges (with $10\%$ increment) of the original graph; and (2) by deleting the same number of edges from the original graph in a targeted way, specifically, among the edges with the largest weights. For each graph $A_i$ we plot the difference between the targeted corruption $T_k(A_i)$ and the random corruption $R_k(A_i)$, for some percentage $k$: $\Delta( W_{q,j}):= \{W_{q,j}(A_i, T_k(A_i))-W_{q,j}(A_i, R_k(A_i))\}_{k=10}^{70}$ against the percentage of deleted edges $i$.

\begin{figure*}[!t]
 \centering
  \subfigure[Shortest-path Distance Metric\label{fig.perturbation.sp}]{
   	\includegraphics[height=80pt]{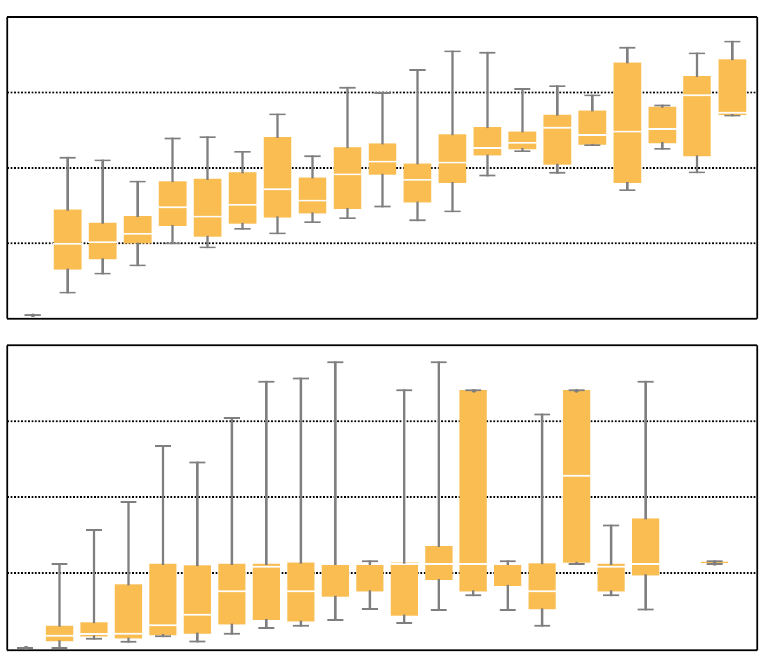}
       \tiny{\put(-24,43){Bottleneck}}
       \tiny{\put(-24,3){Max Norm}}
    \hspace{2pt}
   	\includegraphics[height=80pt]{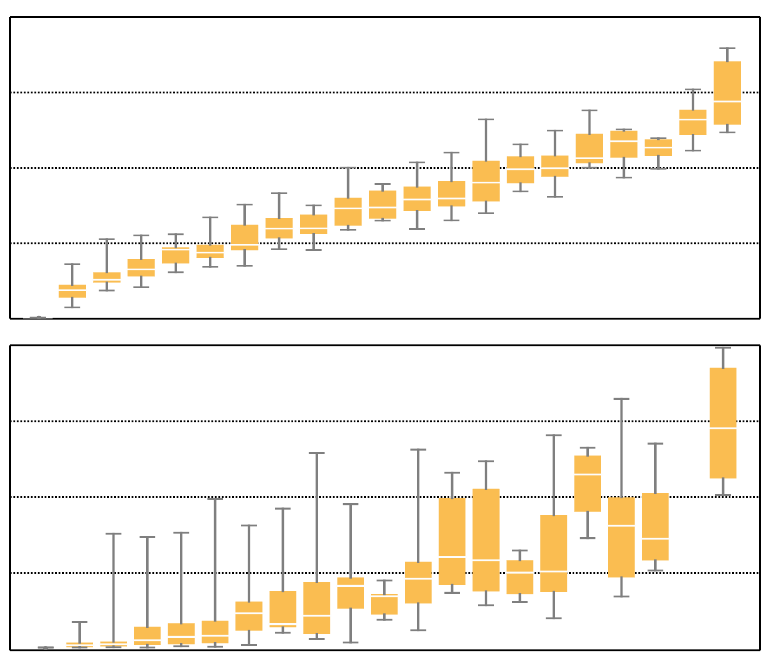}
  } \hspace{15pt}
  \tiny{\put(-27,43){Wasserstein}}
  \tiny{\put(-37,3){Frobenius Norm}}
  \subfigure[Commute-time Distance Metric\label{fig.perturbation.ct}]{
  	\includegraphics[height=80pt]{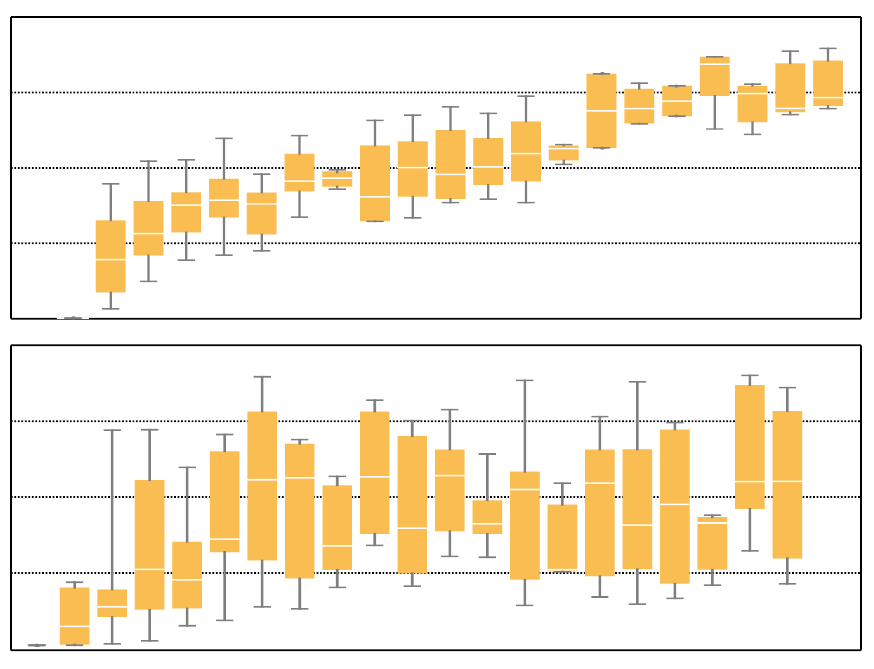} 
    \tiny{\put(-24,43){Bottleneck}}
     \tiny{\put(-24,3){Max Norm}}
    \hspace{2pt}
  	\includegraphics[height=80pt]{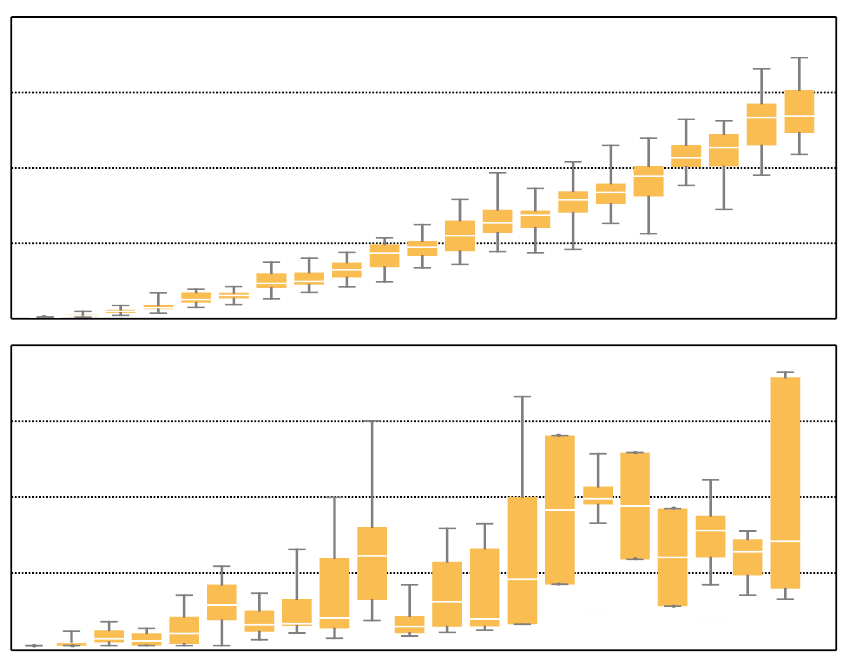}
     \tiny{\put(-27,43){Wasserstein}}
     \tiny{\put(-35,3){Frobenius Norm}}
  }
  \vspace{-2pt}
  \caption{Study of the stability for different similarity measures under small perturbations. The x-axis of each plot shows the percentage of edges deleted from the graph. The y-axis represents the difference between the perturbed graph and the original graph. The y-axes are normalized to $[0,1]$ based upon the maximum observed values.}
  \label{fig.perturbation}
  \vspace{-2pt}
\end{figure*}

We obtain similar observation shown in Figure \ref{fig:P4} as the property 3 test. Our persistence-based similarity measure satisfies the \emph{focus awareness} property in dimension 0 but not in dimension 1. This is due to the fact that the deletion of an edge might create a cycle in the corrupted graph (see the negative values in Figure \ref{fig:P4} bottom).

%% file: sec-discussion-stability.tex
\subsection{Stability Under Perturbation}
\label{sec:stable}

The persistence diagram computation depends on the distance matrix we impose on a graph. A natural question is:  what are the advantages of using the persistence digram on a graph over the distance matrix itself as a topological fingerprint of the graph? 
We would like to give some experimental evidence in this section to justify our choice of persistence-based similarity measure. 

To simplify the analysis, we perturb a small percentage of edges on a simple example, the ``map of science'' graph~\cite{borner2012design} and we focus only on edge deletion.
The experiments we show here only use $PD_0$. 
$PD_1$ is omitted because the results are similar. 
The map of science graph consists of $554$ nodes and $2276$ edges; we refer to it as the \emph{baseline} graph, denoted as $G_0$.

\paragraph*{Edge Deletion Model.}
Our edge deletion is designed as follows. 
For the $i$-th perturbation step, $i\%$ of edges are deleted from the baseline $G_0$ uniformly at random; and such a perturbation is repeated $20$ times to obtain (almost) unbiased results.
We perform a total of $20$ perturbation steps, that is, up to $20\%$ of edges can be deleted from the baseline. 

\paragraph*{Similarity Measures.}
We compare variations among various similarity measure. 
Recall $G_0$ is the baseline graph, and $d_0$ is the distance matrix of its metric space representation. 
Let $G_i$ be an instance of a perturbed graph at the $i$-the perturbation step, $d_i$ be its distance matrix of its metric space representation. 
The first set of similarity measures are based on bottleneck and Wasserstein distances. 
We examine the bottleneck distance $W_\infty$ and the Wasserstein distance $W_2$ between the $0$- and $1$-dimensional persistence diagrams associated with $G_0$ and $G_i$ respectively. 
The second set of similarity measures are based upon matrix norms on the distance matrices. 
We measure the \emph{matrix max norm}, that is, $\left\Vert d_i - d_0\right\Vert_{\rm max}$, where $\left\Vert A\right\Vert_{max} := \max_{ij}|a_{ij}|$ for a matrix $A$. 
We also measure the \emph{matrix Frobenius norm}, that is, 
$\left\Vert d_i - d_0\right\Vert_F$, where $\left\Vert A\right\Vert_{F} : = \sqrt{\sum_i \sum_j (a_{ij})^2}$.  

\paragraph*{Experimental Results.}
Figure~\ref{fig.perturbation} shows our experimental results. 
Figure~\ref{fig.perturbation.sp} uses shortest-path distance metric in the computation of various similarity measures; while Figure~\ref{fig.perturbation.ct} uses commute-time distance metric.

Each subfigure is a box-plot whose y-axis corresponds to a particular similarity measure. 
Since these similarity measures are not directly comparable, the range of y-axis for each plot has been normalized to $[0,1]$ according to the maximum similarity measure across all experimental instances. 

In Figure~\ref{fig.perturbation.sp}, under the shortest-path distance metric, there appears to be a linear relationship between perturbation and the bottleneck distance (and Wasserstein distance). Furthermore, the Wasserstein distance has a smaller variance than the bottleneck distance, making it suitable to study global perturbation in the data. 
On the other hand, similarity measures based on matrix norms are relatively unstable. Both max norm and Frobenius norm show large fluctuations and variance making them less suitable for analysis. 
Moreover, these measures completely fail when the perturbed graph becomes disconnected, which is not an issue for our approach.

Figure~\ref{fig.perturbation.ct}, under the commute-time distance, we observe that persistence-based measure appears to be less noisy and more stable than the shortest-path distance metric.

%% file: sec-conclusion.tex
\section{Conclusion}

Time-varying graphs are becoming increasingly important in data analysis and visualization.
In this paper, we address the problem of capturing and visualizing structural changes in time-varying graphs using techniques originated from topological data analysis, in particular, persistent homology. We provide a simple and intuitive visual interface for investigating structural changes in the graph using persistence-based similarity measures.

There are many on-going and future research avenues based upon our approach. For example, in our work, we restrict topological feature extraction to Rips filtrations. Other types of filtrations, such as clique filtration~\cite{Zomorodian2010}, can be natural in analyzing and understanding time-varying graphs. 

One interesting question that arises in our approach is how best to convert edge weights into distances. The conventional wisdom is that the stronger the communication between nodes (i.e., higher edge weight), the closer together they should be. However, we have some evidence that such a conversion may not always capture the underlying structural changes, and sometimes, an inverse weighting scheme may be more effective. 

It would also be interesting to perform systematic comparison to a wide range of similarity measures in the study of time-varying graphs~\cite{monnig2016resistance}, in particular, to see how these different measures can complement one another in enriching our current visual analytic framework.  

A final note is that we hope the work described here could inspire more graph visualization research to move beyond graph-theoretical measures and venture into techniques from topological data analysis.